\documentclass[11pt]{article}
\usepackage{apacite}
\usepackage{cancel}
 \usepackage{lineno}
\usepackage{amsmath}
\usepackage{amssymb}
\usepackage{graphicx}
\usepackage{xcolor, soul}
\usepackage{enumerate}
\usepackage{enumitem}
\usepackage{verbatim}
\usepackage{hyperref}
\usepackage{url} 
\usepackage{float}
\usepackage{algorithm}
\usepackage{setspace}
\usepackage{algorithm}
\usepackage{algorithmic}
\usepackage{systeme}
\usepackage{mathtools}

\newcommand{\stkout}[1]{\ifmmode\text{\sout{\ensuremath{#1}}}\else\sout{#1}\fi}
\renewenvironment{abstract}
 {\par\noindent\textbf{\abstractname.}\ \ignorespaces}
 {\par\medskip}

\usepackage[tmargin=1in,bmargin=1in,lmargin=1in,rmargin=1in]{geometry}

\usepackage{sectsty} 
\sectionfont{\fontsize{11}{11}\selectfont}
\subsectionfont{\fontsize{11}{11}\selectfont}

\usepackage[normalem]{ulem}
\usepackage{natbib}

\newcommand{\bY}{ {\bf Y} }
\newcommand{\bX}{ {\bf X} }

\newcommand{\bB}{ {\bf B} }

\newcommand{\bS}{ {\bf S} }
\newcommand{\bA}{ {\bf A} }
\newcommand{\bI}{ {\bf I} }
\newcommand{\bH}{ {\bf H} }
\newcommand{\bE}{ {\bf E} }
\newcommand{\bU}{ {\bf U} }

\newcommand{\bZ}{ {\bf Z} }
\newcommand{\bK}{ {\bf K} }
\newcommand{\bC}{ {\bf C} }
\newcommand{\bW}{ {\bf W} }

\newcommand{\bM}{ {\bf M} }
\newcommand{\bG}{ {\bf G} }

\newcommand{\bQ}{ {\bf Q} }
\newcommand{\ba}{ {\bf a} }
\newcommand{\bb}{ {\bf b} }
\newcommand{\bd}{ {\bf d} }

\newcommand{\by}{ {\bf y} }
\newcommand{\bs}{ {\bf s} }
\newcommand{\bx}{ {\bf x} }
\newcommand{\bz}{ {\bf z} }

\newcommand{\bu}{ {\bf u} }

\newcommand{\bzero}{ {\bf 0} }

\newcommand{\bmu}{ \text{\boldmath $ \mu $} }
\newcommand{\bbeta}{ \text{\boldmath $ \beta $} }
\newcommand{\btheta}{ \text{\boldmath $ \theta $} }
\newcommand{\bdelta}{ \text{\boldmath $ \delta $} }

\newcommand{\bSigma}{ \text{\boldmath $ \Sigma $} }
\newcommand{\bLambda}{ \text{\boldmath $ \Lambda $} }
\newcommand{\bPi}{ \mbox{\boldmath $ \Pi $} }

\newcommand{\bTheta}{ \text{\boldmath $ \Theta $} }

\newcommand{\argmin}{\mathop{\mathrm{argmin}}}

\definecolor{maroon}{RGB}{135, 0,25}

\newtheorem{theorem}{Theorem}

\begin{document}

\def\spacingset#1{\renewcommand{\baselinestretch}%
    {#1}\small\normalsize} \spacingset{1}


    \title{{Sparse covariate-driven factorization of high-dimensional\\ brain connectivity with application to site effect correction}}
    \author{Rongqian Zhang$^{1,2}$, Elena Tuzhilina$^1$, Jun Young Park$^{1,3}$\thanks{Corresponding author: \href{mailto:junjy.park@utoronto.ca}{\texttt{junjy.park@utoronto.ca}} }}
   \date{%
    $^1${\small \textit{Department of Statistical Sciences, University of Toronto, Toronto, ON, Canada}}\\%
    $^2${\small \textit{Department of Epidemiology and Biostatistics, University of California San Francisco, San Francisco, CA, United States}}\\%
    $^3${\small \textit{Department of Psychology, University of Toronto, Toronto, ON, Canada}}\\%
}

    \maketitle

\begin{abstract}
{Large-scale neuroimaging studies often collect data from multiple scanners across different sites, where variations in scanners, scanning procedures, and other conditions across sites can introduce artificial site effects. These effects may bias brain connectivity measures, such as functional connectivity (FC), which quantify functional network organization derived from functional magnetic resonance imaging (fMRI). How to leverage high-dimensional network structures to effectively mitigate site effects has yet to be addressed. In this paper, we propose SLACC (Sparse LAtent Covariate-driven Connectome) factorization, a multivariate method that explicitly parameterizes covariate effects in latent subject scores corresponding to sparse rank-1 latent patterns derived from brain connectivity. The proposed method identifies localized site-driven variability within and across brain networks, enabling targeted correction. We develop a penalized Expectation-Maximization (EM) algorithm for parameter estimation, incorporating the Bayesian Information Criterion (BIC) to guide optimization. Extensive simulations validate SLACC’s robustness in recovering the true parameters and underlying connectivity patterns. Applied to the Autism Brain Imaging Data Exchange (ABIDE) dataset, SLACC demonstrates its ability to reduce site effects.}
\end{abstract}


\spacingset{1.2}

\vspace{10mm}

\newpage

\section{Introduction}

Brain magnetic resonance imaging (MRI) offers a powerful tool for studying brain activity {and structure} in both healthy and clinical populations. One of {its} key applications is functional connectivity (FC) derived from functional MRI (fMRI), which quantifies the statistical dependencies between blood-oxygen-level-dependent (BOLD) signals measured over time from two spatially distinct brain regions \citep{biswal1995functional, friston1993functional, greicius2003functional}. FC has been extensively used to investigate brain network organization and its alterations due to neurodevelopment, neuropsychiatric or neurological disease progression or treatment \citep{zonneveld2019patterns,van2010exploring,fox2010clinical}. 

Despite the widespread use of FC, multi-site data collection introduces significant challenges. In large-scale fMRI studies, data collection often involves multiple sites and scanners, which can lead to artificial site effects due to differences in scanner models, scanning protocols, and other local conditions. These effects can  confound analyses and limit the generalizability and reproducibility of findings. Several statistical harmonization methods have been developed to identify and mitigate site effects in neuroimaging data, enabling the construction of harmonized datasets across multiple sites. ComBat, a prominent approach originally proposed for genomics by \citet{johnson2007adjusting}, characterizes site effects as  additive and multiplicative site effects for each feature. It has shown great potential for various neuroimaging data types \citep{fortin2017harmonization, fortin2018harmonization, yu2018statistical}. More recent methods  expand the scope of statistical harmonization to address not only mean-variance differences but also covariance heterogeneity across sites \citep{chen2022mitigating,zhang2023relief, zhang2024san}. In addition, tangent-space harmonization of functional connectivity features using ComBat-GAM has been shown to improve downstream prediction performance in multi-site fMRI studies \citep{zhou2022harmonization}.

Despite these advances, existing harmonization methods for brain connectivity data often treat FC matrices as vectorized features \citep{yu2018statistical, chen2022harmonizing}, thereby ignoring the symmetric network structure in which each element represents the connection strength between two brain regions. This can hinder interpretation of how site effects are identified and corrected within and across brain networks, and may lead to suboptimal harmonization.

In brain network analysis, blind source separation (BSS) and independent component analysis (ICA) have been used in recent methodological advances to decompose FC into low-rank and sparse structures. These structures identify patterns between and within functional networks while quantifying their strength in each subject \citep{eavani2015identifying,cai2017estimation,wang2023locus}. Other studies aim to explore the relationships between connectivity patterns and clinical or demographic factors, as well as other imaging modalities, typically within a linear modeling framework \citep{sun2017store,zhao2021covariate,zhao2022covariance, park2025bayesian}. To reduce the number of parameters, it is common to impose low-rank structures either on the FC matrix itself \citep{zhao2021covariate} or on the regression parameters \citep{sun2017store}. However, as \citet{zhao2021covariate} and \citet{park2025bayesian} noted, incorporating the high dimensionality of the whole-brain network remains a challenge.
 
We hypothesize that leveraging recent advances in latent variable modeling, particularly in low-rank matrix/tensor factorization, would be beneficial for improving the harmonization of brain connectivity data. Building on this, we propose a covariate-driven latent factor model for brain connectivity, which decomposes connectivity data into a linear combination of low-rank, sparse connectivity patterns, weighted by covariate factors. The sparsity assumption is motivated by two considerations. First, the human brain is organized into distinct functional networks, each involving a relatively small number of regions \citep{salvador2005neurophysiological}. Sparse connectivity patterns provide a parsimonious representation of these localized connections between brain regions \citep{eavani2015identifying}. Second, because connectivity matrices are high-dimensional, enforcing sparsity in latent patterns reduces the effective number of parameters, resulting in more stable and reliable estimates while preventing the risk of spurious findings. 

By embedding site and covariate effects into these latent factors, our model aims to identify and correct localized site-specific heterogeneity within and across brain networks, while also accounting for biological effects from covariates. Our model achieves a more efficient and flexible representation of multi-site connectivity matrices by explicitly using site effects to drive the decomposition, avoiding the need for vectorization and thus preserving the network properties and the dependence structure of the connectivity matrix. Moreover, incorporating low-rank structure and regularization into modeling the sparse connectivity patterns reduces model complexity. Together, these efforts not only allow for the accurate localization and correction of site effects but also improve the interpretability of these effects in multi-site connectivity data. 

The rest of this paper is organized as follows. In Section \ref{sec:method}, we introduce SLACC (Sparse LAtent Covariate-driven Connectome) factorization, a low-rank sparse latent factor model for decomposing and harmonizing multi-site connectivity matrices, along with a penalized Expectation-Maximization (EM) algorithm. In Section \ref{sec:sim}, we conduct extensive simulations to demonstrate SLACC's ability to recover true parameters and underlying connectivity patterns. In Section \ref{sec:data}, we apply our method to the resting-state fMRI data from the Autism Brain Imaging Data Exchange (ABIDE) study and compare its performance with other harmonization approaches. We conclude with a discussion of the proposed method in Section \ref{sec:dis}.

\section{Methods}\label{sec:method}

\subsection{Notation and setup}

In this paper, we formulate and present our model under a multi-site setting for generality, although a single-site setting can be considered as a special case. Let $i=1,\dots, M$ denote the site index, $j=1,\dots, n_i$ denote the subject index within the $i$th site, and $n=\sum_{i=1}^M n_i$ denote the total number of subjects. Let $\bx_{ij}\in\mathbb{R}^q$ be the covariate vector for subject $j$ in site $i$, and let $\bX\in\mathbb{R}^{n\times q}$ be the matrix obtained by stacking the $\bx_{ij}$'s in rows across sites and subjects. Let $\bY_{ij}$ be a $V\times V$ brain connectivity matrix for subject $j$ in site $i$, where $V$ denotes the number of brain regions. We define $\by_{ij}=\mathcal{T}(\bY_{ij})$ to be the vector of $p=V(V+1)/2$ upper-triangular entries of $\bY_{ij}$, including the diagonal entries; $\mathcal{T}(\cdot)$ is defined similarly for any symmetric matrix input.
$\bA\odot\bB$ refers to the Hadamard product of $\bA$ and $\bB$ of the same dimension, and $\ba\circ\bb$ refers to the outer product of $\ba$ and $\bb$. $\text{diag}(\ba)$ refers to the diagonal matrix with entries $\ba$. $||\ba||_q$ and $||\bA||_F$ refer to the  $L_q$ norm of $\ba$ and Frobenius norm of $\bA$, respectively.

\subsection{Model} \label{sec:model}
\begin{figure}[h] 
         \centering      \includegraphics[scale=0.6]{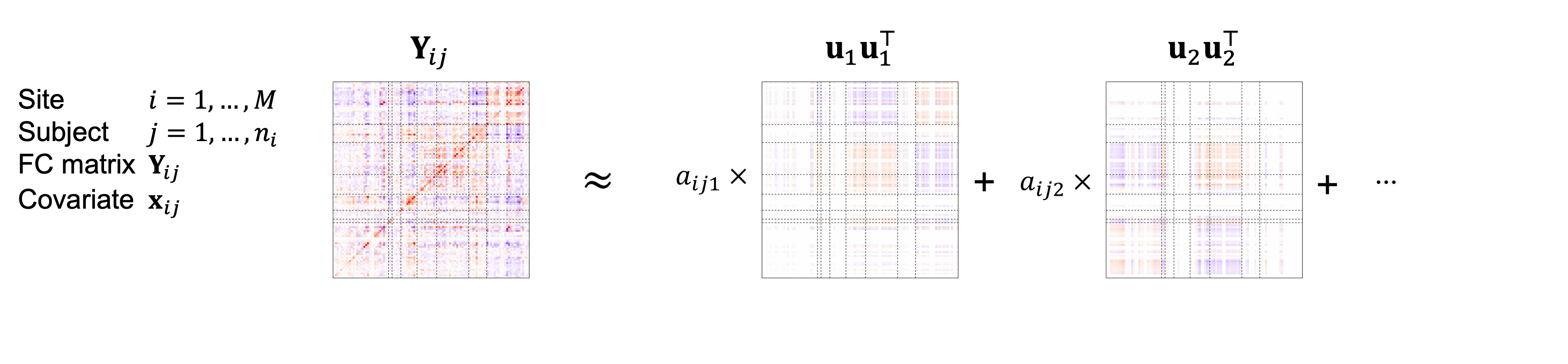}    
         \vspace{-5mm}\caption{Illustration  of  framework decomposing a subject's  connectivity matrix into sparse rank-1
connectivity patterns $\{\bu_l\bu_l^\top\}_{l=1}^L$, weighted by subject scores $a_{ijl}$ modeled by covariates $\bx_{ij}$.}\label{diagblock}
\end{figure}

Our model separates the connectivity matrix $\bY_{ij}$ into shared rank-1 latent connectivity patterns and subject-specific scores. Covariate/site effects act on the subject scores, so that covariates explain systematic biological variation, while site effects are allowed to influence both the mean and variance of the latent scores as well as the residual variation. To formalize this, we state
\begin{align}\label{eq_1}
\bY_{ij}&=\sum_{l=1}^L a_{ijl} \times \bu_l\bu_l^\top+\bE_{ij}.
\end{align}
Here, each $\bu_l=(u_{l1},\dots,u_{lV})^\top \in \mathbb{R}^V$ represents a vector of region weights for the $l$th latent pattern, and $L$ represents the number of latent connectivity patterns. Also, we let
\begin{align*}
  \bE_{ij}[v,v^\prime]\overset{i.i.d}{\sim}  \mathcal{N}(0,\phi_{i}^2),\quad 1\leq v\le v^\prime\leq V.  
\end{align*}
Note that \eqref{eq_1} is a common signal-plus-noise model to decompose and factorize brain connectivity, a specification widely used in the fMRI literature \citep{eavani2015identifying, zhao2021covariate, wang2023locus}, and the rank-1 patterns $\bu_l\bu_l^\top$ are motivated  by the observation that FC patterns often exhibit block-diagonal structure, as illustrated in Figure \ref{diagblock}. 

We  assume a linear model for $a_{ijl}$ to incorporate covariate/site effects:
\begin{align}\label{eq_2} 
a_{ijl} = \bx_{ij}^\top\bbeta_{l}+\delta_{ijl},
\end{align}
with $\delta_{ijl}\overset{i.i.d}{\sim}\mathcal{N}(0,\sigma_{il}^2)$, $j=1,\dots,n_i$. Also, we assume that $\bE_{ij}$ is independent of $\delta_{ijl}$ for all $i,j,l$. In our model, the site effects are characterized by three sources: site-specific mean effects and covariate effects (via $\bx_{ij}^\top\bbeta_l$), site-specific latent-factor variances ($\sigma_{il}^2$), and site-specific noise variance ($\phi_i^2$), providing flexibility in capturing site-specific heterogeneity in mean and covariance. 

We note that a latent factor need not be strongly covariate-driven; for example, \(\bu_l\) may still be recovered when \(\bbeta_l=0\) and \(\sigma^2_{1l}=\cdots=\sigma^2_{Ml}\), provided that its latent variance is large. Also, the linear model for the scores is  a parsimonious  specification and could be extended to longitudinal or additive models, although such extensions would require more complex estimation procedures.

\subsection{Likelihood}
Under \eqref{eq_1}–\eqref{eq_2}, $\bY_{ij}$ follows a multilinear normal distribution, and we derive its likelihood  of  $\by_{ij}=\mathcal{T}(\bY_{ij})$. We define $\bs_l=\mathcal{T}(\bu_l\bu_l^\top) \in \mathbb{R}^p$, and let $\bS = [\bs_1;\dots;\bs_L] \in \mathbb{R}^{p\times L}$ and $\bB = [\bbeta_1;\dots;\bbeta_L] \in \mathbb{R}^{q\times L}$. Let $\bTheta = \lbrace\bB, \bU, \sigma_{il}^2, \phi_i^2; i = 1,\dots,M, l=1,\dots,L\rbrace$ be the parameter set. Then $\by_{ij}$ follows a multivariate normal distribution with
\begin{align*}
    \bmu_{ij}
&= \text{E}(\by_{ij}| \bx_{ij})
= \sum_{l=1}^L (\bx_{ij}^\top \bbeta_l)\,\bs_l
= \bS \bB^\top \bx_{ij},\\
\bSigma_i
&= \mathrm{Cov}(\by_{ij}|\bx_{ij})
= \sum_{l=1}^L \sigma_{il}^2 \bs_l \bs_l^\top + \phi_i^2 \bI_p
= \bS \text{diag}(\sigma_{i1}^2, \dots, \sigma_{iL}^2) \bS^\top + \phi_i^2 \bI_p,
\end{align*}
where $\bI_p$ is a $p\times p$ identity matrix.
Then, the negative log–likelihood  can then be written as
\begin{align}\label{eq_ll}
    -{\log\mathcal{L}}(\bTheta|\by_{ij})=\frac{1}{2}\log(2\pi) +\frac{1}{pn}\left\lbrace\sum_{i=1}^M\frac{n_i}{2}\log|\bSigma_i|+\sum_{i=1}^M\sum_ {j=1}^{n_i} \frac{1}{2}(\by_{ij}-\bmu_{ij})^\top\bSigma_i^{-1}(\by_{ij}-\bmu_{ij})\right\rbrace.
\end{align}

\subsection{Identifiability}\label{sec:identifiability}

Let $\mathbb{Y}$ and $\mathbb{E}$ denote the $n \times V \times V$ tensors  formed by stacking $\bY_{ij}$ and $\bE_{ij}$ with the first mode corresponding to subjects, and the second and third modes corresponding to  regions.  Then, under the 3-way CANDECOMP/PARAFAC (CP) decomposition,
\begin{align}\label{eq_kway}
  \mathbb{Y}= \sum_{l=1}^L \ba_l \circ \bu_l \circ \bu_l+\mathbb{E}, \quad \quad  {\ba_l=\bX\bbeta_l+\bdelta_l,}
\end{align} 
where $\bdelta_l=(\delta_{11l},\dots, \delta_{1n_1 l}, \dots, \delta_{M1l},\dots, \delta_{Mn_Ml})^\top$. Under the CP-decomposed form, we obtain the following result that equivalence of the likelihood implies equivalence of the parameters. The proof follows naturally from the result of \citet{lock2018supervised}.

\begin{theorem}\label{thm_1} 
The parameter set $\bTheta$ is identifiable under the following regularity conditions:
\begin{enumerate}
\item[A.1] $\min(n, q, L)$ $\geq$ 2 and $L<V$.
\item[A.2] {$\bU=[\bu_1;\dots; \bu_L]$ is of full column rank}, and  $||\bu_l||_2=1$ for $l=1,\dots,L$.
\item[A.3] The first nonzero entries of $\bu_l$ for $l=1,\dots,L$ are positive.
    \item[A.4] $\sigma^2_{11}\ge\sigma^2_{12}\ge\dots\ge\sigma^2_{1L}>0$.
    \item[A.5] $\bX$ is of full column rank.
\end{enumerate}

 \end{theorem}
A.1 is needed to meet Kruskal's uniqueness condition that is required to establish identifiability, and A.2-A.4 are needed to ensure identifiability under scaling, sign flipping, and re-ordering. 

\subsection{Parameter estimation}\label{alg}

As the objective function in \eqref{eq_ll} is not convex with respect to $\bTheta$, we first develop a penalized EM algorithm  for fixed $L$ in this section, then discuss the choice of $L$ in Section \ref{rank}.

\subsubsection{E-step}\label{Estep1}

Treating $\ba_{ij}\equiv(a_{ij1},\dots,a_{ijL})^{\top}$ as latent variables, the E-step  updates $\ba_{ij}$ by its conditional expectation given the data $\by_{ij}$ and the parameter estimates from the previous step:
\begin{align}\label{E_a}
 \text{E}(\ba_{ij}|\by_{ij})  &=\bSigma_{\ba_{ij}|\by_{ij}}\{\bSigma^{-1}_{\ba_{ij}}(\bB^\top\bx_{ij})+\bS^\top\bSigma^{-1}_{\by_{ij}|\ba_{ij}}\by_{ij}\},
\end{align}
where $\bSigma_{\ba_{ij}}=\text{diag}(\sigma_{i1}^2,\dots, \sigma_{iL}^2),$ $\bSigma_{\by_{ij}|\ba_{ij}}=\phi_i^2\cdot \bI_p$, and $\bSigma^{-1}_{\ba_{ij}|\by_{ij}}=\bS^\top\bSigma^{-1}_{\by_{ij}|\ba_{ij}}\bS+\bSigma^{-1}_{\ba_{ij}}.$

\subsubsection{M-step: Updating $\bbeta_l$, $\sigma_{il}^2$, and $\phi_i^2$} \label{Mstep1}

We maximize the conditional expectation of the joint log-likelihood of $\ba_{ij}$ and $\by_{ij}$. Since the joint log-likelihood of $\by_{ij}$ and $\ba_{ij}$ can be decomposed into the log-likelihood of $\ba_{ij}$ and the conditional log-likelihood of $\by_{ij}$ given $\ba_{ij}$, the M-step is partitioned into two optimization problems:
\begin{align}\label{eq_m1}
\max_{\lbrace\bbeta_l\rbrace_{l=1}^L, \{\sigma_{il}^2\}_{i=1,l=1}^{M,L}} \quad 
\sum_{i=1}^M\sum_{j=1}^{n_i}\text{E}_{\ba_{ij}|\by_{ij}}[\log\mathcal{L}(\ba_{ij})]
\quad\text{and}\quad
\max_{\{\bu_l\}_{l=1}^L,\{\phi_i^2\}_{i=1}^M} \quad 
\sum_{i=1}^M\sum_{j=1}^{n_i}\text{E}_{\ba_{ij}|\by_{ij}}[\log\mathcal{L}(\by_{ij}|\ba_{ij})].
\end{align}
For  $\{\bbeta_l,\sigma_{il}^2,\phi_i^2\}$, the closed-form solutions can be derived in the M-step. However, updating $\bu_l$ requires an additional  procedure that we address in Section \ref{optimizeu}. The first optimization problem in  \eqref{eq_m1} can be further separated into two parts, one updating $\bbeta_l$ and the other  updating $\sigma_{il}^2$. 

\paragraph{Updating $\bbeta_l$.}
The part of the expected complete-data log-likelihood that depends on $\bbeta_l$ is
\begin{align*}
Q(\bbeta_l)
= -\frac12\sum_{i=1}^M\sum_{j=1}^{n_i} \frac{(a_{ijl}-\bx_{ij}^\top\bbeta_l)^2}{\sigma_{il}^2} +\text{constant},    
\end{align*}
which is reduced to the weighted least squares problem, yielding the closed form as follows:
\begin{align}\label{updateB}
    \widehat{\bbeta}_l = \left(\sum_{i=1}^M\sum_{j=1}^{n_i} \frac{\bx_{ij} \bx_{ij}^\top}{\sigma_{il}^2}\right)^{-1} \left( \sum_{i=1}^M\sum_{j=1}^{n_i}\frac{\bx_{ij} a_{ijl}}{\sigma_{il}^2}  \right).
\end{align}

\paragraph{Updating $\sigma_{il}^2$.}

The site-specific conditional second-moment estimator is provided by 
\begin{align}\label{updateD}
\frac{1}{n_i}\sum_{j=1}^{n_i}\{
(\ba_{ij}-\bB^\top\bx_{ij}) 
(\ba_{ij}-\bB^\top\bx_{ij})^\top + \bSigma_{\ba_{ij}|\by_{ij}}\}.    
\end{align}
From here, $\lbrace\hat{\sigma}_{il}^2\rbrace_{i=1}^M$ is obtained by taking diagonal entries of the above form. 

\paragraph{Updating $\phi_i^2$.}

$\phi_i^2$ is updated by
\begin{align} \label{updatePhi}
\hat{\phi}_i^{2}
&=\frac{1}{n_ip}\sum_{j=1}^{n_i}\text{E}_{\ba_{ij}|\by_{ij}}||\by_{ij}-{\bS}\ba_{ij}||_2^2=\frac{1}{n_ip}\sum_{j=1}^{n_i}||\by_{ij}-{\bS}\ba_{ij}||_2^2
+\frac{1}{p}\,\mathrm{tr}(\bS\,\bSigma_{\ba_{ij}|\by_{ij}}\,\bS^\top).
\end{align}

\subsubsection{M-step: updating $\bu_l$ with regularization}\label{optimizeu}

Given the parameter updates in Sections \ref{Estep1} and \ref{Mstep1}, the M-step for updating $\bU\equiv [\bu_1;\dots;\bu_L]$ without  penalization is equivalent to 
\begin{align} \label{eq:unpenalized}
   \widehat{\bU}=\underset{\bU}{\argmin}\sum_{i=1}^M\sum_{j=1}^{n_i}\dfrac{||\bY_{ij}-\bU{\bPi}_{ij}\bU^\top||_F^2+ \text{tr}((\bU^\top\bU \odot \bU^\top\bU)\bSigma_{\ba_{ij}|\by_{ij}})}{4\phi_i^{2}} ,
\end{align}
where  $\bPi_{ij}=\text{diag}(a_{ij1},\dots, a_{ijL})\in\mathbb{R}^{L\times L}$. Its derivation is presented in Section \ref{supp:secA} of the supplementary material. To achieve sparsity, we propose using $L_0$ regularization on $\bU$  for two reasons. First, it provides estimates with less bias compared with other regularization methods (e.g., Lasso and ridge penalties), while still ensuring  sparsity. This step is critical for harmonization, as it allows for more accurate identification and correction of site effects. Second, $L_0$ regularization enables us to incorporate the Bayesian Information Criterion (BIC) directly into the M-step, thus avoiding the need for computationally expensive fine-tuning of parameters. 

When all the parameters other than $\bU$ are fixed, the penalized objective guided by BIC is:
\begin{align}   \label{BIC2}
   \widehat{\bU}=\underset{\bU}{\argmin} \sum_{i=1}^M\sum_{j=1}^{n_i}\dfrac{||\bY_{ij}-\bU{\bPi}_{ij}\bU^\top||_F^2+ \text{tr}((\bU^\top\bU \odot \bU^\top\bU)\bSigma_{\ba_{ij}|\by_{ij}})}{2\phi_i^{2}}+\log(n) \cdot ||\bU||_0.
\end{align}
Solving  \eqref{BIC2} poses two computational challenges related to non-convexity. First, directly optimizing the $L_0$ penalty is computationally intractable. To mitigate this issue, we consider the Truncated Lasso Penalty (TLP), a surrogate for $L_0$ penalty expressed by $||\bU||_0 \approx \sum_{v=1}^V\sum_{l=1}^L \min \left(\frac{|u_{vl}|}{\tau},1\right)$  \citep{shen2012likelihood}. Here, $\tau>0$ is a tuning parameter controlling the degree of approximation, and smaller values of $\tau$ makes TLP close to the $L_0$ penalty. Following the  implementation of  \citet{shen2012likelihood} where the default $\tau$ depends on the dimensionality, we choose $\tau=0.5\cdot \sqrt{\log(VL)/n}$ as a default. Second, even the (unpenalized) objective in $\bU$ contains the bilinear form $\bU\bPi_{ij}\bU^\top$, making  \eqref{BIC2} computationally challenging. We handle this by using the Alternating Direction Method of Multipliers (ADMM) splitting, which mitigates the non-convexity by alternating convex subproblems. Specifically, we introduce an auxiliary copy $\bU^\star$ of $\bU$  and enforce the consensus constraint $\bU=\bU^{\star}$. 

Addressing the challenges above results in the updated objective
\begin{align}\label{updateU}
\widehat{\bU}=\underset{\bU}{\argmin} &\sum_{i=1}^M\sum_{j=1}^{n_i}\dfrac{||\bY_{ij}-\bU^\star{\bPi}_{ij}\bU^\top||_F^2+ \text{tr}((\bU^{\star \top}\bU \odot \bU^{\star\top}\bU)\bSigma_{\ba_{ij}|\by_{ij}})}{2\phi_i^{2}} \nonumber\\
&+\frac{\log(n)}{2} \sum_{v=1}^V\sum_{l=1}^L \dfrac{I(|u_{vl}|\leq \tau) }{\tau} |u_{vl}| + \frac{\log(n)}{2} \sum_{v=1}^V\sum_{l=1}^L \dfrac{I(|u_{vl}^{\star}|\leq \tau) }{\tau} |u_{vl}^\star|, 
\end{align}
subject to $\bU=\bU^\star$, which  we solve  using ADMM by alternating updates of $\bU$ and $\bU^\star$ under the consensus constraint $\bU=\bU^\star$. Analogous to the one-step update  in \citet{shen2012likelihood}, we replace $I(|u_{v,l}|\le \tau)$ and $I(|u_{v,l}^{\star}|\le \tau)$  with the estimates from the previous step, i.e., $I(|u_{v,l}^{(t-1)}|\le \tau)$ and $I(|u_{v,l}^{\star(t-1)}|\le \tau)$. The details are provided in Section  \ref{supp:secB} of the supplementary material.

\subsection{Algorithmic considerations}\label{sec:algoconsiderations}

Our algorithm for fixed $L$ is summarized in Algorithm \ref{alg:EM}. Since each update of $\bU$ in the penalized EM algorithm could lead to a local minimum, especially when $L$ is large, we improve algorithmic stability by first implementing the unpenalized EM algorithm (i.e., solving \eqref{eq:unpenalized}) initialized with higher-order singular value decomposition \citep{de2000multilinear}, and then using the resulting estimates to initialize the penalized EM algorithm. We compare different initialization methods in Section \ref{sec:sensitivity} of the supplementary material. Also, we find that an annealing strategy that gradually increases the penalty in Equation \eqref{updateU} from 0 (unpenalized EM) to $\log(n)$ at each iteration of the EM mitigates the local minimum problem.

\begin{algorithm}[htbp]
\caption{Penalized EM algorithm, when $L$ is provided}\label{alg:EM}
\begin{algorithmic}[1]
    \STATE \textbf{Initialize:} Initialize $\lbrace\bU^{(0)}$, $\ba_{ij}^{(0)}, \bB^{(0)}, \sigma_{il}^{2(0)}$, $\phi_i^{2(0)}\mid i=1,\dots, M, l=1,\dots,L\rbrace$.\\
    \FOR{$t = 1, 2, \dots$ until convergence}
     \STATE \textbf{E-step}: Given $\bU^{(t-1)}$,$ \bB^{(t-1)}$, $\sigma_{il}^{2(t-1)}$,and $\phi_i^{2(t-1)}$, update $\ba_{ij}^{(t)}$ using  \eqref{E_a}.
     \STATE \textbf{M-step}: Given $ \ba_{ij}^{(t)}$, $\phi_{i}^{2(t-1)}$, and $\bU^{(t-1)}$ update $\bU^{(t)}$ by solving  \eqref{updateU}. 
     \STATE \textbf{M-step}: Given $ \ba_{ij}^{(t)}$, $\bU^{(t)}$, $\sigma_{il}^{2(t-1)}$ and $\phi_i^{(t-1)}$ update $ \bB^{(t)}$ and  $\sigma_{il}^{2(t)}$,  using  \eqref{updateB} and \eqref{updateD}.
     \STATE \textbf{M-step}: Given $\ba_{ij}^{(t)}$, $ \bU^{(t)}$, $\phi_i^{(t-1)}$ and $\sigma_{il}^{2(t)}$, update $\phi_i^{2(t)}$ using  \eqref{updatePhi}.
             \STATE \textbf{Check for convergence:} If $ ||\bU^{(t)}-\bU^{(t-1)}||_F < \epsilon$, stop the iteration. 
  \ENDFOR
\STATE \textbf{Normalize}: Normalize $\bu_l^{(t)}$ with $||\bu_{l}^{(t)}||_2=1$ and make corresponding adjustments to $\bB$, ${\sigma_{il}^2}$.
  
    \STATE \textbf{Return:} $\widehat{\bTheta} =\lbrace\bU^{(t)},\bB^{(t)},\sigma_{il}^{2(t)}, \phi_i^{2(t)}\mid i=1,\dots, M, l=1,\dots,L\rbrace$
\end{algorithmic}
\end{algorithm}

\subsection{Selection of $L$}\label{rank}

Up to this point, we have assumed that $L$ in our model is known. To choose $L$ from the data, we consider the (extended) BIC \citep{chen2008extended}, which is defined as: 
\begin{align}
    \text{EBIC}({\widehat{\bTheta}})=-2\sum_{i=1}^M\sum_{j=1}^{n_i}{\log\mathcal{L}}(\widehat\bTheta|\by_{ij}) +\log(n)\cdot \text{df}(\widehat{\bTheta}) + 2\gamma \log(p) \cdot \text{df}(\widehat{\bTheta}), \label{eq:BICL}
\end{align}
where $\text{df}(\widehat{\bTheta})=qL+ML+M+||\widehat\bU||_0$,  the number of nonzero elements of the parameter set. $0\leq\gamma\leq 1$ is to be chosen by user, where $\gamma=0$ corresponds to the usual BIC, and higher values of $\gamma$ (e.g., $\gamma=0.5$) could be used if $p$ is deemed to be high. More data-driven methods may be used, such as site-stratified cross-validation, as illustrated in Section \ref{sec:cv} of the supplementary material.

\subsection{Harmonization}\label{sec:harmonization}

After estimating parameters, we construct the harmonized data by removing site-specific means, normalizing site-specific variances for each latent factor, and rescaling the residual variances. Note that, if $\bx_{ij}$ includes biological covariates $\bz_{ij}$ and site dummy variables, then $\bx_{ij}^\top\bbeta_l$ can be rewritten as $\alpha_l + \bz_{ij}^\top \btheta_l + \gamma_{il}$ with the constraint $\sum_{i=1}^M \gamma_{il}=0$. Since the site-specific heterogeneity is embedded in both the latent factors and the residual noise, we harmonize them as
\begin{align}\label{harmo}
a_{ijl}^{(h)}=\frac{\sigma_{l}^{(h)}}{\hat\sigma_{il}}(\hat a_{ijl}-\hat\alpha_l-\bz_{ij}^\top\hat\btheta_{l}-\hat\gamma_{il})+\hat\alpha_l+\bz_{ij}^\top\hat\btheta_{l},\quad \bE_{ij}^{(h)}=\frac{\phi^{(h)}}{\hat\phi_i}\widehat\bE_{ij},
\end{align}
where $\sigma_{l}^{(h)}=\sqrt{\frac{\sum_{i=1}^Mn_i\hat\sigma_{il}^2}{n}}, \phi^{(h)}=\sqrt{\frac{\sum_{i=1}^Mn_i\hat\phi_{i}^2}{n}}$. The harmonized data is therefore given by
\begin{align*}
    \bY_{ij}^{(h)}=\sum_{l=1}^L a_{ijl}^{(h)}{\hat\bu_l\hat\bu_l^\top}+\bE_{ij}^{(h)}.
\end{align*}

\section{Simulation studies}\label{sec:sim}

\subsection{Simulation 1}

\begin{figure}[h]
    \centering
    \includegraphics[width=0.5\linewidth]{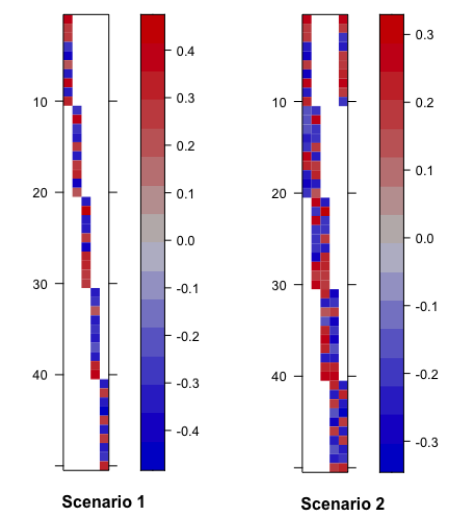}
    \caption{True $\bU\in\mathbb{R}^{50\times 5}$ for scenarios 1 and 2 used in simulation studies.}
    \label{fig:U}
\end{figure}

Simulation 1 evaluates parameter estimation under varying sample sizes ($n=100, 200, \dots, 500$), when true $L$ is specified correctly. We specify $M=2, L=5, V=50$ $(p=1275)$, and $q=4$. We considered two possible scenarios of $\bU$ as illustrated in Figure \ref{fig:U}: (i) Scenario 1 is the case where there are no overlapping entries across $V$ and 20\% of the entries are nonzero, and scenario 2 has overlapping elements and 40\% of the entries are nonzero. Each nonzero entry for both scenarios was generated once from Uniform(0.5, 1.0) and multiplied by either 1 or -1 randomly, then each column of $\bU$ was normalized to have unit norm. We used 4 covariates in total, consisting of two continuous biological covariates and two site-specific intercepts. The biological covariates were generated from $\mathcal{N}(0,1)$ and the corresponding regression coefficients were also generated from $\mathcal{N}(0, 1)$ and kept fixed throughout the simulation. Site-specific intercepts for each $l$ were fixed to be 0.3 for site 1 and -0.3 for site 2 for every $l=1, \dots, 5$. For variance components, we set $\sigma_{11}^2=1, \sigma_{11}^2=2, \dots, \sigma_{15}^2=5$ for site 1 and $\sigma_{21}^2=5, \sigma_{22}^2=4, \dots, \sigma_{25}^2=1$ for site 2. Lastly, we set  $\phi_1^2=1.2$ and $\phi_2^2=0.8$. It yielded the median of $\sigma^2$ parameters divided by the median of $\phi^2$ parameters equal to 3, which is considerably lower than the fit from the real data analysis in Section \ref{sec:data} in which the ratio was nearly 25.  For each sample size, we repeated the simulation $B=1000$ times. After fitting the model, the column order and signs of $\widehat{\bU}$ were aligned with $\bU$ based on their correlations, and the other parameters were reordered accordingly.

Due to the lack of statistical methods addressing the same problem, we considered two related methods as competitors: (i) (\textbf{SLACC-True}) SLACC with true $\bU$ provided, which serves as a benchmark, and (ii) (\textbf{SLACC-NoPen}) SLACC without penalization in the M-step in updating $\bU$. To compare the performances, we use the following metrics. We first compare mean square error (MSE) for $\bB,  \lbrace \sigma_{il}^2 \rbrace_{i=1,l=1}^{2,5}, \lbrace\phi_i^2\rbrace_{i=1}^2$. For example, the MSE for $\widehat{\bU}$ is computed by $1/B\cdot \sum_{b=1}^{B} ||\widehat{\bU}^{(b)} - \bU||_F^2$ where $\widehat{\bU}^{(b)}$ is the estimated $\bU$ from the $b$th simulation. The variance ($1/B\cdot \sum_{b=1}^{B} ||\widehat{\bU}^{(b)} - \sum_{b=1}^{B}\bU^{(b)}/B||_F^2$) as well as bias$^2$ (defined as  MSE$-$variance) are reported. Second, we evaluated sensitivity and specificity of $\widehat{\bU}$. Sensitivity is defined as the proportion that $\widehat{\bU}^{(b)}$ captures the true nonzero indices out of the number of nonzero entries of $\bU$, and specificity is defined as the proportion that $\widehat{\bU}^{(b)}$ captures the true zero indices out of the number of zero entries of $\bU$. Note that, trivially,  the sensitivity and specificity of SLACC-True are both 100\% and the values for SLACC-NoPen are 100\% and 0\% respectively.

\begin{figure}[!] 
         \centering      
         \includegraphics[width=0.9\linewidth]{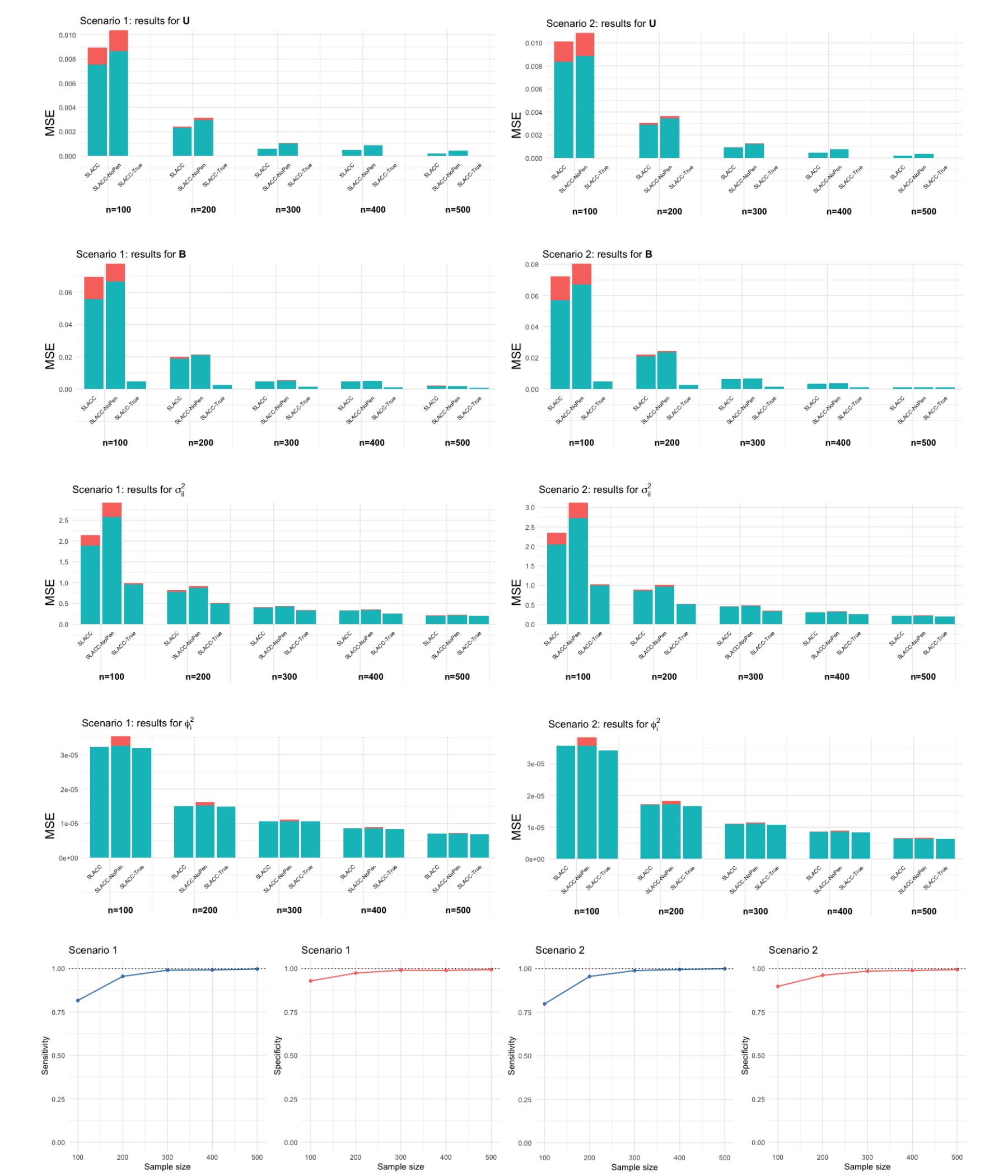}
        \caption{Results of Simulation 1. The left and right columns describe scenarios 1 and 2, respectively. The first 4 rows describe MSE (red+blue), bias$^2$ (red) and variance (blue). The last row shows sensitivity and specificity of the estimated $\bU$.
}\label{com_sim}
\end{figure}

Figure \ref{com_sim} shows the results in Simulation 1. The overall patterns look similar in both scenarios, and we see a clearly decreasing MSE as sample size increases bounded above the performance of SLACC-true in all parameters. SLACC achieved consistently lower MSE than SLACC-NoPen in estimating $\bU$, while the differences in performance for the other parameters became marginal as the sample size increased. We also note that, attributed to the surrogate $L_0$ penalization, the bias in SLACC's parameter estimation was similar to SLACC-NoPen or often lower (e.g., in estimating $\phi_i^2$). Also, the surrogate $L_0$ penalty in estimating $\bU$ achieved high sensitivity and specificity in all $n$ except for $n=100$, which imply the case where the low signal-to-residual-noise (e.g., $\bB$ and $\sigma_{il}^2$ in relative to $\phi_i^2$) impedes the estimation of $\bU$.

\subsection{Simulation 2} \label{sec:bicsim}

Simulation 2 evaluates possible methods in selecting the true $L=5$ used in Simulation 1. We fit the model with varying $L$ with $L\in \lbrace 2,3, \dots, 10\rbrace$ and extracted BIC ($\gamma=0$ in \eqref{eq:BICL}), EBIC $(\gamma=0.5)$, and 5-fold cross-validation loss for each fit. The results of estimated $L$ over 200 simulations are reported in Figures \ref{fig:simulation2S1} and \ref{fig:simulation2S2} in Section \ref{supp:secC} of the supplementary material. As expected, all these methods closely chose the true $L$ as $n$ increases. BIC was the most promising in our simulation designs, while EBIC was conservative in small samples and cross-validation was slightly more anti-conservative than BIC. Compared to scenario 1 where $\bu_l$ were orthogonal and the nonzero indices were not overlapping, a higher $n$ was needed to consistently recover the true $L$ as shown in Scenario 2.

\section{Data analysis}\label{sec:data}
\subsection{Data preparation and preprocessing}

We use preprocessed resting-state fMRI (rs-fMRI) data from the Autism Brain Imaging Data Exchange (ABIDE) \citep{di2014autism} to empirically evaluate our model's performance. ABIDE is a multi-site consortium providing rs-fMRI data from 408 subjects with Autism Spectrum Disorder (ASD) and 476 controls. Due to the lack of prior coordination between sites, scan parameters, scanner types, and diagnostic/assessment protocols varied across sites. The sample size for each site ranged from 21 (CALTECH) to 169 (NYU). In addition to the existing evidence of site effects on FC \citep{yu2018statistical}, we also note that the number of scans per person varied substantially across sites, ranging from 78 (OHSU) to 296  time points (UM), which serves as design-based evidence of site effects in variances because FC is a statistical estimate whose variance depends on the number of time points. 

We used the ABIDE dataset preprocessed using the Configurable Pipeline for the Analysis of Connectomes (CPAC), which includes slice-timing correction, motion realignment, and intensity normalization. Further preprocessing details are provided in \citet{craddock2013towards}. The fMRI time-series for a set of regions of interest (ROIs) were obtained from  the Harvard-Oxford (HO) atlas \citep{desikan2006automated,frazier2005structural}, which consists of $V=110$ ROIs that we subsequently mapped to Yeo's 7-network parcellation \citep{yeo2011organization} plus the subcortical network. The resulting 8 functional networks include the Visual Network (VN), Subcortical Network (Scor), Somatomotor Network (SMN),  Salience/Ventral Attention Network (SN/VAN),  Limbic Network (LN), Frontoparietal Control Network (FPCN), Dorsal Attention Network (DAN), and Default Mode Network (DMN). 

We excluded 38 subjects with incomplete brain coverage and the CMU site  due to its limited sample size (5 subjects). The final sample comprises 845 subjects across 16 sites. We then computed Fisher-transformed Pearson correlation matrices from the  fMRI time-series data so that the variance depends only on the number of time points.  Since our goal is to understand the individual differences in connectivity, we removed the sample mean across subjects for each edge as a preprocessing step. The diagonal entries were set to 0 and were not used in estimating $\phi_i^2$.

\subsection{Results}

\begin{figure}[h]
    \centering
    \includegraphics[width=0.55\linewidth]{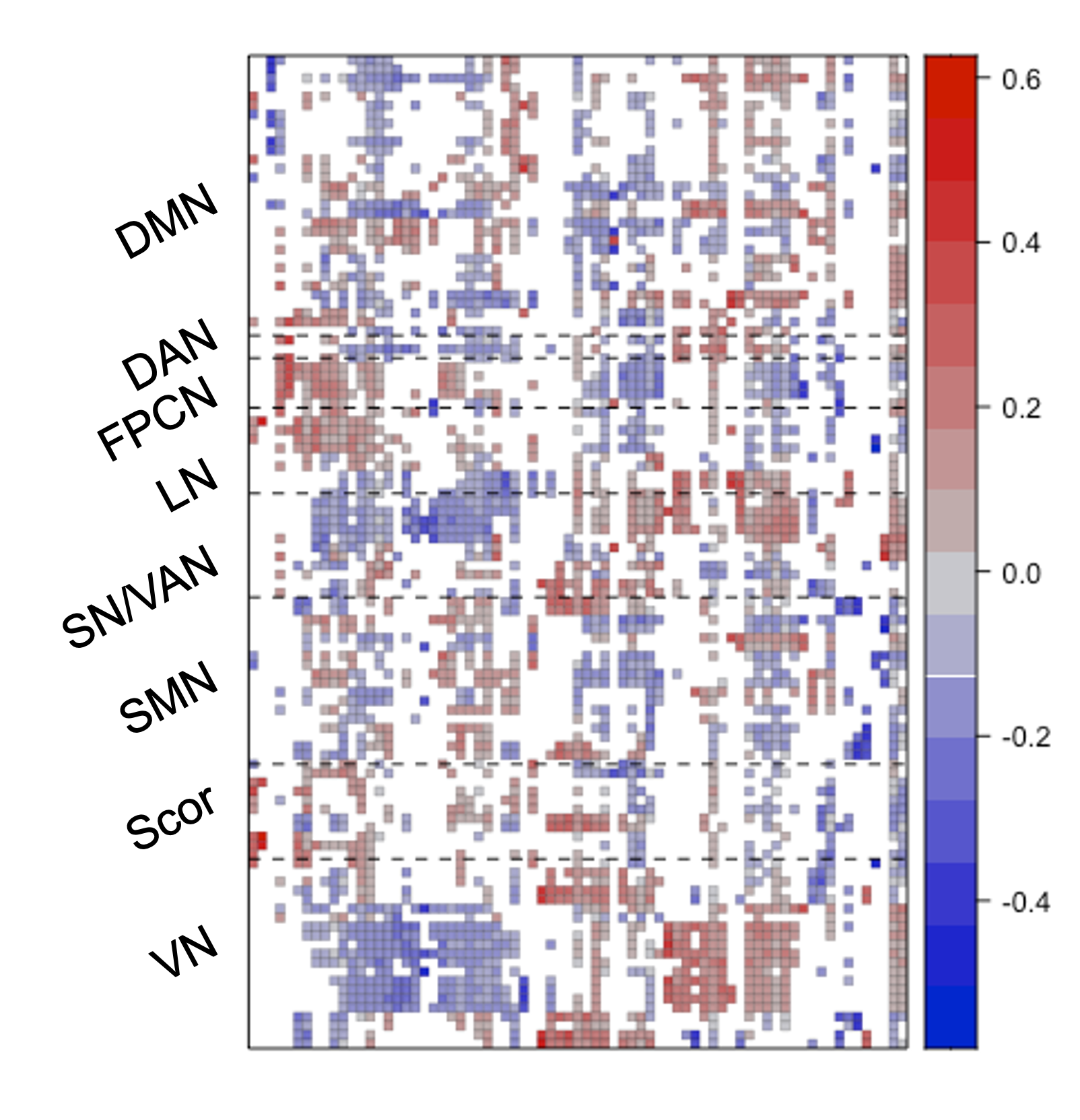}
    \caption{Estimated ${\bU}$ ($L=73$) using the training data ($n=591$), with the column orders obtained by hierarchical clustering.}
    \label{fig:Uest}
\end{figure}

We considered age, $\text{age}^2$, sex, diagnosis, age$\times$sex, and age$\times$diagnosis as biological covariates to allow for nonlinear age effects and age-dependent heterogeneity by sex and diagnosis while keeping the model parsimonious \citep{sanders2023age,alaerts2016sex}. We split the data into two folds consisting of 591 (70\%) training and 254 (30\%) test subjects, stratified by  study site. We fit our model to the training data, with $L=73$ selected by BIC. About 68.2\% of the entries in $\widehat{\bU}$ were 0 (Figure \ref{fig:Uest}). Our analysis reveals that the covariate-dependent latent patterns appear to vary across within-network and between-network structures. For example, a subset of regions in visual network (VN) had noticeable similarity in latent patterns, and these appeared to be related to the salience/ventral attention network (SN/VAN) as well as the frontoparietal control network (FPCN). At the same time, some latent patterns were very sparse and concentrated on a very few regions not clearly aligned with network structure.

We used the estimated parameters to extract the latent subject scores $\hat{a}_{ijl}$ for both the training and test data. Then, using the estimated parameters, we further constructed the `batch-free' model with the parameters $\lbrace\hat{\bu}_l, \hat{\btheta}_l, \hat{\alpha}_l, \sigma_{l}^{(h)}, \phi^{(h)}\rbrace_{l=1}^L$ defined in Section \ref{sec:harmonization}, and obtained the corrected latent subject scores, denoted by $\hat{a}_{ijl}^{\star}$, for both training and test data. We calculated two $F$ test statistics for each edge to test (i) equality of means across sites and (ii) equality of variances across sites using absolute residuals as in Levene's test \citep{carroll1985note}.

\begin{figure}[h]
    \centering
    \includegraphics[width=0.75\linewidth]{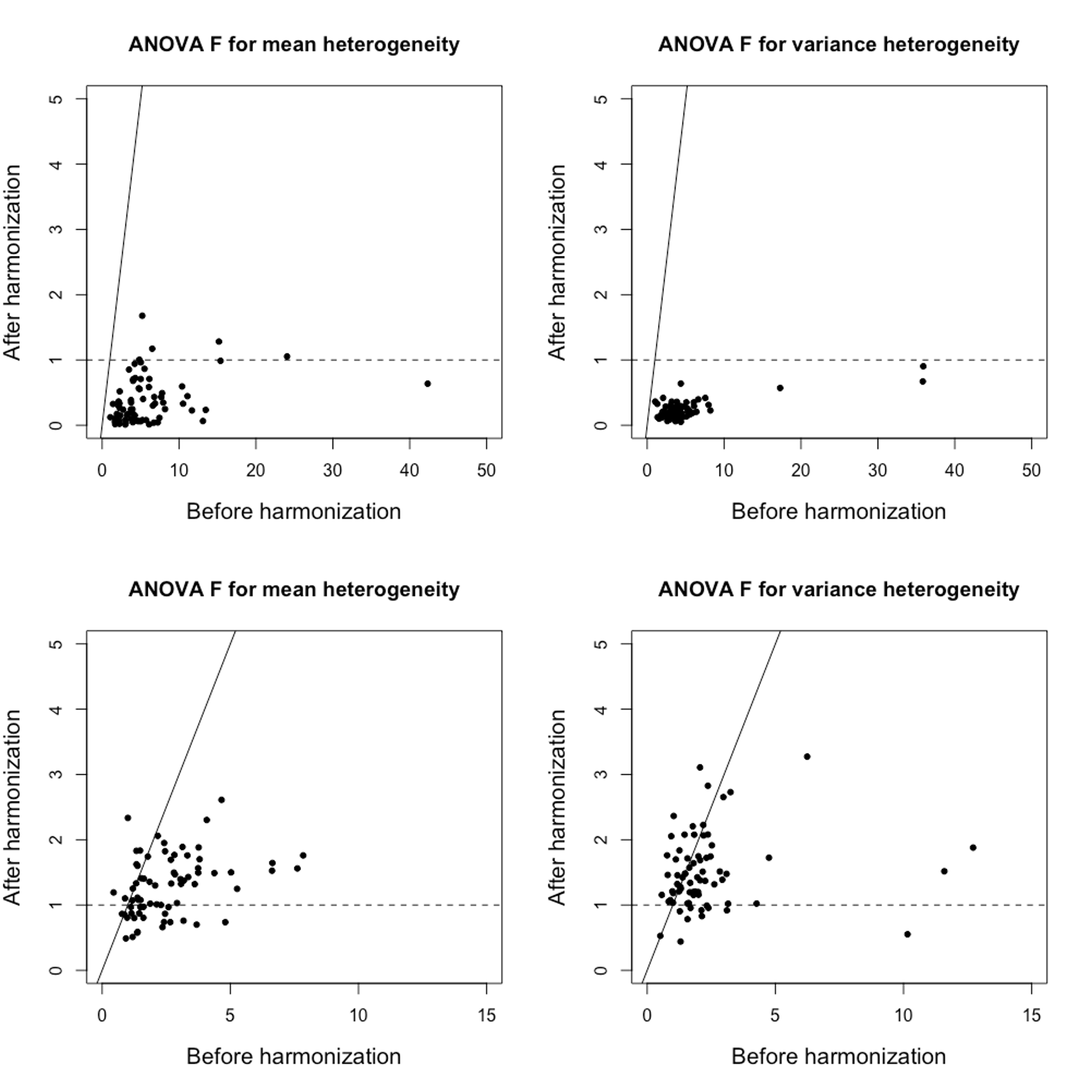}
    \caption{$F$ test statistic for measuring mean and variance heterogeneity across sites, using both training (first row) and test data (second row). The dotted horizontal line is the expected value of the null $F$ distribution (i.e., no site effects). The solid line is the 45 degree line.}
    \label{fig:abide_result}
\end{figure}

Because only a very minor portion of the variability in the latent scores was explained by biological covariates, whereas site effects were more prominent, we focus on analyzing site effects before and after harmonization. Based on Figure \ref{fig:abide_result}, SLACC identified latent factors in the training data that are highly associated with site effects in means and variances, and a few of them exhibited large effects (i.e., $F>10$ in the training data).  After harmonization, most statistics shrank toward the null level. Although an average $F$ statistic below 1 (the expected value of the $F$ distribution under the null) may occur due to overfitting to the training data, the test data also show that most $F$ statistics were reduced toward the nominal level.

We  compared the edge-level harmonization performance of SLACC with ComBat. For ComBat, we used \texttt{ComBatFamily} (\url{https://github.com/andy1764/ComBatFamily}, ver. 0.2.1) to fit the model on the training data and  harmonize the test data. More advanced methods, such as out-of-sample CovBat harmonization, are not currently supported in \texttt{ComBatFamily} and would require additional considerations regarding the number of retained principal components relative to the test sample size and potential overfitting. We therefore did not include them as primary comparators. At the edge level, the median $F$ statistics for mean and variance were 1.40 and 1.08 for SLACC, compared with 1.25 and 1.15 for ComBat. This suggests that SLACC more effectively harmonizes variances, whereas ComBat more effectively harmonizes means. In addition, we evaluated how well our method preserves biological variation. For each harmonized test data, we first used the MDMR test with Euclidean distance and 10000 permutations \citep{shehzad2014multivariate} to evaluate multivariate associations between FC and covariates. For age, both the linear and quadratic terms were statistically significant at $\alpha<0.05$ ($p$s$<$0.0001). For ComBat, diagnosis was also significant ($p=0.028$). For the edge-wise analysis, we extracted $p$ values for each edge; ComBat had 7.58\% of the $p$ values less than 0.05, and SLACC had 7.40\%. While ComBat was slightly more effective in preserving biological variation, we also note that biological covariates including age and sex were correlated with sites in the ABIDE data, which might have confounded the harmonization step, and such exploration would require additional investigation \citep{bridgeford2025no}.

Lastly, to address possible issues related to positive semidefiniteness, we conducted an additional analysis. Following \citet{desai2025connectivity}, we projected each subject's FC data onto the Fisher space using a matrix-log transformation, and then applied SLACC to the off-diagonal entries. The results are reported in Section \ref{sec:datasupp} of the supplementary material. We found that, in the Fisher space, site effects in the $L=41$ latent scores selected by BIC were much more pronounced than those in the element-wise Fisher-transformed correlations, with median $F=343.62$ and 9.95 for mean and variance heterogeneity, respectively, while the groupings of $\widehat{\bU}$ did not reveal noticeable network-specific patterns. In the element-wise analysis, SLACC had lower $F$ statistics on the test data for both means (median $F=1.10$ for SLACC and 1.22 for ComBat) and variances (median $F=1.09$ for SLACC and 1.15 for ComBat).

\section{Discussion}\label{sec:dis}

In this paper, we proposed SLACC, a structured latent factor model for  brain connectivity that represents connectivity matrices through sparse rank-1 patterns and covariate-dependent subject scores. When noticeable site effects are present, the proposed framework provides an interpretable way to characterize site effects through latent mean, latent variance, and residual variance components, while leveraging low-rank and sparse structure to regularize estimation in high dimensions.

Our work connects recent advances in brain connectivity factorization with the problem of statistical harmonization. Compared with existing approaches for sparse blind source separation or covariate-assisted connectivity modeling, SLACC places these ideas in a generative framework that accommodates covariate effects and admits identifiable parameterization under suitable conditions. The rank-1 restriction is central to this formulation: it enables parameter identifiability, direct sparsity regularization on the latent patterns, and flexible modeling of subject scores, although it may also limit the ability to capture more complex connectivity structures.

The proposed model has some room for extension. First, we assume that the vectorized connectivity data follow a multivariate normal distribution. This Gaussian formulation is computationally convenient because it yields a tractable penalized EM algorithm with  closed-form updates, and it is also reasonable in our fMRI application, where the observed connectivity measures are Fisher-transformed correlations. However, this distributional assumption may be less appropriate for structural connectivity derived from diffusion tensor imaging, whose entries are known to be  sparse and zero-inflated.  Second, the Gaussian model on vectorized connectivity does not guarantee positive semidefinite harmonized correlation matrices. This is less problematic for edge-wise BWAS analyses and for non-correlation connectivity measures, but future work could incorporate matrix-log, Wishart, or other correlation-matrix-valued formulations. Such adaptations, however, may compromise the computational efficiency (e.g., closed-form solutions in estimation) and interpretability (e.g., mean, variance, and covariance specifications of site effects) offered by the current normal distribution framework. Third, SLACC decomposes connectivity matrices into a set of rank-1 latent pattern matrices, which may limit the method's ability to recover neural circuits with complex topological structures or spatial patterns that extend beyond a rank-1 structure \citep{wang2023locus}. A potential extension of the SLACC framework could involve adopting more flexible higher-rank representations to better capture such complexity. Nevertheless, this extension would require addressing challenges such as ensuring identifiability. Fourth, uncertainty quantification and hypothesis testing for model parameters in latent subject scores are not straightforward, and they might require conditioning on the $\hat{\bu}_l$s being known, as \citet{zhao2022covariance} also noted.

\section*{Software}

The R package for SLACC is  available at  \url{https://github.com/junjypark/SLACC}. Scripts reproducing results in Section \ref{sec:sim} are available at \url{https://github.com/junjypark/SLACC_reproduce}.

\section*{Acknowledgments}

We would like to thank the editor, associate editor, and reviewers for providing insightful suggestions, which  significantly improved the quality of this manuscript. We would like to thank Dr. Piotr Zwiernik (University of Toronto/Universitat Pompeu Fabra) and Dr. Lindsay D. Oliver (Centre for Addiction and Mental Health) for their helpful suggestions.

\section*{Funding}
RZ was partially supported by the University of Toronto’s Data Sciences Institute through the doctoral student fellowship. ET was partially supported by the Natural Sciences and Engineering Research Council of Canada (NSERC) under grant RGPIN-2023-04727, the University of Toronto Data Science Institute, and the University of Toronto McLaughlin Centre.
JYP was partially supported by the NSERC under grant RGPIN-2022-04831, the University of Toronto Data Science Institute, and the Connaught Fund. The computing resources were enabled in part by support provided by University of Toronto and the
Digital Research Alliance of Canada.

\section*{Conflict of interest}
None declared.

\spacingset{1}

\bibliographystyle{apacite} 
\bibliography{refs.bib}

\newpage
\appendix
\spacingset{1.2}

\renewcommand{\thefigure}{S\arabic{figure}}

\setcounter{figure}{0}

\begin{center}
    {\Large Supplementary materials}
\end{center}

\section{Derivation of Equation \eqref{eq:unpenalized}} \label{supp:secA}

Recall our model is summarized by 
$\bY_{ij}=\bU\bPi_{ij}\bU^\top+\bE_{ij}$, with  
$\bPi_{ij}=\mathrm{diag}(a_{ij1},\dots,a_{ijL})$ and 
$\mathcal{T}(\bE_{ij})\sim \mathcal{N}\!\left(\bzero,\phi_i^2\bI_p\right)$.
Up to an additive constant, the conditional negative log-likelihood is
\begin{align}
-\log p(\bY_{ij}|\ba_{ij};\bU)
~\propto~
\frac{1}{2\phi_i^2}||
\mathcal{T}\!(\bY_{ij}-\bU\bPi_{ij}\bU^\top)
||_2^2 .
\label{eq:SX_vecnll}
\end{align}
Here, we optimize a surrogate form using a Frobenius norm: $\left||
\mathcal{T}(\bM)
\right||_2^2
~=~
\kappa\cdot
\left||
\bM
\right||_F^2$, where $\kappa$ is a constant that depends only on the choice of $\mathcal{T}(\cdot)$. Since $\kappa$ does not depend on $\bU$, it does not affect the optimizer. Equation \eqref{eq:SX_vecnll} is equivalent (up to constants) to
\begin{align}
-\log p(\bY_{ij}| \ba_{ij};\bU)
~\propto~
\frac{1}{4\phi_i^2}||
\bY_{ij}-\bU\bPi_{ij}\bU^\top
||_F^2.
\label{eq:SX_fronll}
\end{align}

As a preliminary step, let $\hat{\ba}_{ij}=\text{E}(\ba_{ij}|\by_{ij})$ and 
$\bQ_{ij}=\mathrm{Cov}(\ba_{ij}|\by_{ij})$ be the posterior mean and covariance from the E-step.
Write
\begin{align*}
\ba_{ij}=\hat{\ba}_{ij}+\bd_{ij},\qquad 
\text{E}(\bd_{ij}|\by_{ij})=\bzero,\qquad 
\text{E}(\bd_{ij}\bd_{ij}^\top| \by_{ij})=\bQ_{ij}.
\end{align*}
Define $\widehat{\bPi}_{ij}=\mathrm{diag}(\hat{a}_{ij1},\dots,\hat{a}_{ijL})$ so that
$\bPi_{ij}=\widehat{\bPi}_{ij}+\mathrm{diag}(\bd_{ij})$.
Then, consider the conditional expectation of the squared Frobenius loss in Equation \eqref{eq:SX_fronll}:
\begin{align}
\text{E}[
||\bY_{ij}-\bU\bPi_{ij}\bU^\top||_F^2
~|~ \by_{ij}
]
&=
\text{E}[
||\bY_{ij}-\bU\widehat{\bPi}_{ij}\bU^\top-\bU\mathrm{diag}(\bd_{ij})\bU^\top||_F^2
~|~ \by_{ij}
]\nonumber\\
&=
||\bY_{ij}-\bU\widehat{\bPi}_{ij}\bU^\top||_F^2
+\text{E}[||\bU\mathrm{diag}(\bd_{ij})\bU^\top||_F^2~|~\by_{ij}].
\label{eq:SX_split}
\end{align}
Note that the cross-term vanishes because $\text{E}(\bd_{ij}| \by_{ij})=\bzero$.

Now, it is sufficient to obtain the second term of Equation \eqref{eq:SX_split}. Using
$\bU\mathrm{diag}(\bd_{ij})\bU^\top=\sum_{l=1}^L d_{ijl}\,\bu_l\bu_l^\top$,
we obtain
\begin{align*}
||\bU\mathrm{diag}(\bd_{ij})\bU^\top||_F^2
=
\sum_{l=1}^L\sum_{l'=1}^L d_{ijl}d_{ijl'}\,(\bu_l^\top\bu_{l'})^2
=
\bd_{ij}^\top(\bU^\top\bU\odot\bU^\top\bU)\bd_{ij}.
\end{align*}
Taking expectation and using
$\text{E}(\bd_{ij}\bd_{ij}^\top| \by_{ij})=\bQ_{ij}$ yields
\begin{align}
\text{E}[||\bU\mathrm{diag}(\bd_{ij})\bU^\top||_F^2~|~\by_{ij}]
=
\mathrm{tr}((\bU^\top\bU\odot\bU^\top\bU)\,\bQ_{ij}).
\label{eq:SX_trace}
\end{align}
Combining \eqref{eq:SX_fronll}, \eqref{eq:SX_split}, and \eqref{eq:SX_trace}, the M-step objective for updating $\bU$ reduces to Equation \eqref{eq:unpenalized}. 

\newpage
\section{Details of the ADMM algorithm for  $\bU$ in the penalized M-step}\label{supp:secB}

\subsection{Penalized M-step objective for updating $\bU$ via Equation \eqref{updateU}}

 Let $\bQ_{ij}=\bQ_i=\bSigma_{\ba_{ij}| \by_{ij}}\in\mathbb{R}^{L\times L}$ denote the conditional covariance of $\ba_{ij}$ in the E-step.
In our implementation, $\bQ_i$ is site-specific and does not depend on $j$. Define weights $w_{ij}=1/\phi_i^2$ and $w_{\text{sum}}=\sum_{i=1}^M\sum_{j=1}^{n_i} w_{ij}$.
Then the penalized M-step for $\bU$ (after one-step linearization of the TLP penalty) can be written as
\begin{align}
\min_{\bU,\bU^\star}\quad
&\sum_{i=1}^M\sum_{j=1}^{n_i}\frac{w_{ij}}{2}
\lbrace
||\bY_{ij}-\bU^\star\bPi_{ij}\bU^{\top}||_F^2
+\text{tr}((\bU^{\star\top}\bU\odot\bU^{\star\top}\bU)\bQ_i)
\rbrace
\nonumber\\
&\quad + \lambda ||\bC^{(t-1)}\odot \bU||_1
      + \lambda ||\bC^{\star (t-1)}\odot \bU^\star||_1
\qquad \text{subject to}\qquad \bU=\bU^\star,
\label{eq:S1_obj}
\end{align}
where $\lambda=\log(n)/2$ in our paper. Following Equation \eqref{updateU}, the $(v,l)$th entry of $\bC^{(t-1)}$ are $\frac{1}{\tau}\,I(|u^{(t-1)}_{vl}|\le \tau)$, and the entries for $\bC^{\star (t-1)}$ are defined accordingly.

\subsection{ADMM splitting and augmented Lagrangian}

To handle the nonsmooth $L_1$ terms and the symmetry constraint $\bU=\bU^\star$, we introduce auxiliary variables
$\bZ$ and $\bZ^\star$ and impose
$
\bU=\bZ, \bU^\star=\bZ^\star,$ and $\bU=\bU^\star
$.
Let $f(\bU,\bU^\star)$ denote the smooth part of \eqref{eq:S1_obj}.
We solve
\begin{align*}
\min_{\bU,\bU^\star,\bZ,\bZ^\star}\quad
&f(\bU,\bU^\star)+\lambda||\bC^{(t-1)}\odot\bZ||_1+\lambda||\bC^{\star(t-1)}\odot\bZ^\star||_1 \nonumber\\
\text{such that }\quad
&\bU=\bZ,\qquad \bU^\star=\bZ^\star,\qquad \bU=\bU^\star.
\end{align*}
Using the scaled ADMM form, we introduce dual variables $\bW,\bW^\star$ for $(\bU-\bZ)$ and $(\bU^\star-\bZ^\star)$,
and $\bLambda$ for $(\bU-\bU^\star)$.
With penalty parameters $\rho>0$ and $\eta>0$, the scaled augmented Lagrangian becomes
\begin{align}
\mathcal{L}_{\rho,\eta}
=& f(\bU,\bU^\star)
+\lambda||\bC^{(t-1)}\odot\bZ||_1
+\lambda||\bC^{\star(t-1)}\odot\bZ^\star||_1 \nonumber\\
&+\frac{\rho}{2}||\bU-\bZ+\bW||_F^2
+\frac{\rho}{2}||\bU^\star-\bZ^\star+\bW^\star||_F^2
+\frac{\eta}{2}||\bU-\bU^\star+\bLambda||_F^2.
\label{eq:S2_augLag}
\end{align}

\subsection{Closed-form updates for $\bU$ and $\bU^\star$}

At iteration $k$, updating $\bU$ reduces to solving
\begin{align}
\bU^{(k+1)}
=\argmin_{\bU}
f(\bU,\bU^{\star (k)})
+\frac{\rho}{2}||\bU-\bZ^{(k)}+\bW^{(k)}||_F^2
+\frac{\eta}{2}||\bU-\bU^{\star (k)}+\bLambda^{(k)}||_F^2.
\label{eq:S3_Usubprob}
\end{align}
Let $\bG_{U^\star}$ and $\bH_{U^\star}$ be defined by
\begin{align*}
\bG_{U^\star} &= \sum_{i=1}^M\sum_{j=1}^{n_i} w_{ij}\,\bPi_{ij}\,(\bU^{\star\top}\bU^\star)\,\bPi_{ij} \quad \quad \text{and} \quad \quad
\bH_{U^\star} = \sum_{i=1}^M\sum_{j=1}^{n_i} w_{ij}\,\bY_{ij}\,\bU^\star\,\bPi_{ij},
\end{align*}
and define the aggregated conditional second-moment matrix $\overline{\bQ}=\frac{1}{w_{\text{sum}}}\sum_{i=1}^M\sum_{j=1}^{n_i}w_{ij}\bQ_i$. Then the optimality condition for \eqref{eq:S3_Usubprob} yields the linear system
\begin{align*}
\bU^{(k+1)}\,\bK_{U^\star}
=
\bH_{U^\star}
+\rho(\bZ^{(k)}-\bW^{(k)})
+\eta(\bU^{\star (k)}-\bLambda^{(k)}),
\end{align*}
where
\begin{align*}
\bK_{U^\star}
=
\bG_{U^\star}
+(\rho+\eta)\bI_L
+w_{\text{sum}}\,(\bU^{\star (k)\top}\bU^{\star (k)}\odot \overline{\bQ}).
\end{align*}
Therefore,
\begin{align*}
\bU^{(k+1)}
=(\bH_{U^\star}+\rho(\bZ^{(k)}-\bW^{(k)})+\eta(\bU^{\star (k)}-\bLambda^{(k)}))\bK_{U^\star}^{-1}.
\end{align*}
The $\bU^\star$-update is analogous with $\bU$ and $\bU^\star$ swapped.

\subsection{Soft-thresholding updates and dual steps}

Updating $\bZ$ reduces to the following objective:
\[
\bZ^{(k+1)}
=\argmin_{\bZ}
\lambda||\bC^{(t-1)}\odot\bZ||_1
+\frac{\rho}{2}||\bU^{(k+1)}-\bZ+\bW^{(k)}||_F^2,
\]
which has the elementwise soft-thresholding solution
\begin{align*}
Z^{(k+1)}_{vl}
=\mathcal{S}_{(\lambda/\rho)C^{(t-1)}_{vl}}\!(U^{(k+1)}_{vl}+W^{(k)}_{vl}),
\qquad
\mathcal{S}_{\kappa}(x)=\mathrm{sign}(x)\cdot \max(|x|-\kappa,0).
\end{align*}
The update for $\bZ^\star$ is identical with $(\bU^\star,\bW^\star,\bC^{\star(t-1)})$.

The dual variables are updated by
\begin{align*}
\bW^{(k+1)} &= \bW^{(k)}+\bU^{(k+1)}-\bZ^{(k+1)},\\
\bW^{\star (k+1)} &= \bW^{\star (k)}+\bU^{\star (k+1)}-\bZ^{\star (k+1)},\\
\bLambda^{(k+1)} &= \bLambda^{(k)}+ \bU^{(k+1)}-\bU^{\star (k+1)}.
\end{align*}

\newpage

\section{Sensitivity analysis on the initialization methods}\label{sec:sensitivity}

We consider several alternative methods for initializing $\bU$. In the current algorithm, we first (i) obtain an initial estimate of $\bU$ using HOSVD and then (ii) run SLACC without penalization to produce the initialization of $\bU$ used in the penalized EM algorithm. To assess the sensitivity of the procedure to the choice of the initial value in step (i), we replace HOSVD with several alternative initialization methods while keeping step (ii) unchanged, and evaluate the resulting unpenalized log-likelihood in Equation \eqref{eq_ll} of the main manuscript. The methods considered are as follows:

\begin{enumerate}[nolistsep]

    \item (Proposed) Fit a HOSVD to $\mathbb{Y}$ in Section \ref{sec:identifiability} and use the resulting estimate to initialize $\bU$.

    \item Fit a CP decomposition to $\mathbb{Y}$ in Section \ref{sec:identifiability} and use the resulting estimate to initialize $\bU$.

    \item Use a random initialization of $\bU$, where each entry is independently generated from a standard normal distribution and each column is normalized so that $||\bu_l||_2 = 1$ for $l = 1, \dots, L$.

\end{enumerate}
As a gold standard, we also considered using the true $\bU$ to initialize, which serves as a gold standard.

\begin{figure}[h]
    \centering
    \includegraphics[width=0.85\linewidth]{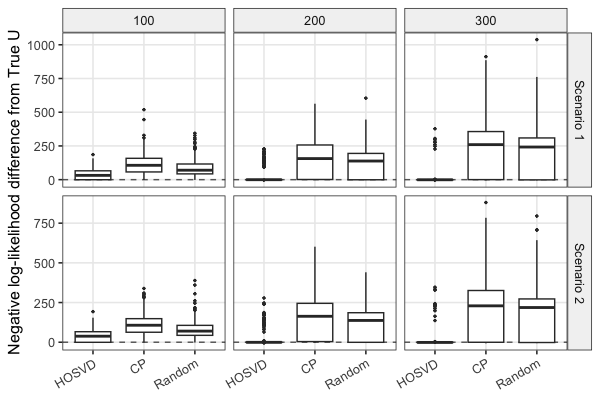}
    \caption{Comparison of initialization methods relative to the gold-standard initialization.}
    \label{fig:HOSVD_Sensitivity}
\end{figure}

We used Simulation 1 from Section \ref{sec:sim} with $n\in\lbrace 100, 200, 300\rbrace$, and each simulation was repeated 500 times. We computed the difference in negative log-likelihood between each method and the gold standard, and the boxplots of the results are summarized in Figure \ref{fig:HOSVD_Sensitivity}. We see that HOSVD-based initialization achieves performance closest to the gold standard across all sample sizes considered, making it the preferred choice.

\newpage
\section{Site-stratified cross validation to choose $L$}\label{sec:cv}

We implement site-stratified cross-validation as follows:

\begin{enumerate}[nolistsep]

    \item For each site \(i=1,\dots,M\), randomly divide the \(n_i\) subjects into \(K\) approximately equal-sized folds. The folds are then combined across sites to form \(K\) site-stratified validation splits. For a given split, let \(\mathcal{I}_{\mathrm{test},i}\) and \(\mathcal{I}_{\mathrm{train},i}\) denote the validation and training indices, respectively, for site \(i\).

    \item For each candidate \(L\),

    \begin{enumerate}[label=(\alph*)]

        \item Fit SLACC on the training data and obtain \(\widehat{\bTheta}_{\mathrm{train}}^{(L)}\).

        \item Compute the validation loss

        \[
        \sum_{i=1}^M \sum_{j\in \mathcal{I}_{\mathrm{test},i}}
        -\log f\!(\by_{ij}\mid \bx_{ij}; \widehat{\bTheta}_{\mathrm{train}}^{(L)}).
        \]

        \item Repeat over all \(K\) folds and average the resulting validation losses.

    \end{enumerate}

    \item Choose \(L\) to minimize the average validation loss.

    \end{enumerate}

\newpage
\section{Supplementary figures for Simulation 2 results}\label{supp:secC}

\begin{figure}[h]
    \centering
    \includegraphics[width=0.3\linewidth]{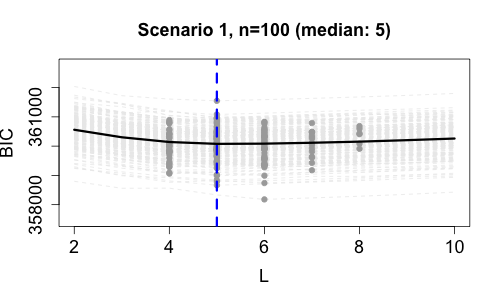}
    \includegraphics[width=0.3\linewidth]{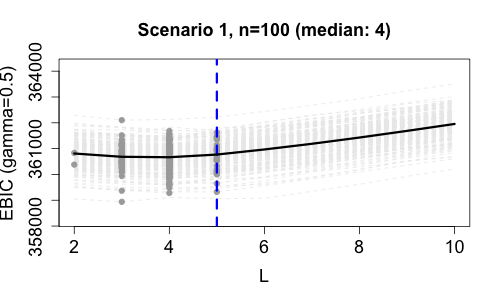}
    \includegraphics[width=0.3\linewidth]{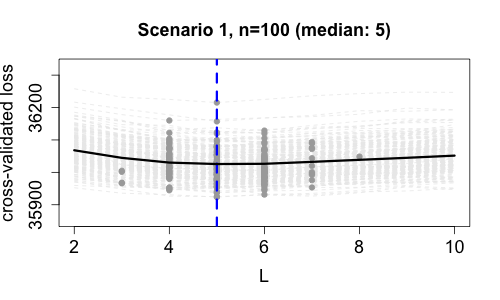}
    
    \includegraphics[width=0.3\linewidth]{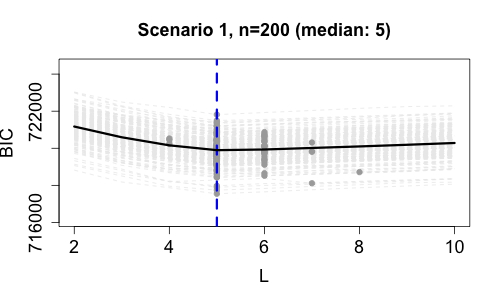}
    \includegraphics[width=0.3\linewidth]{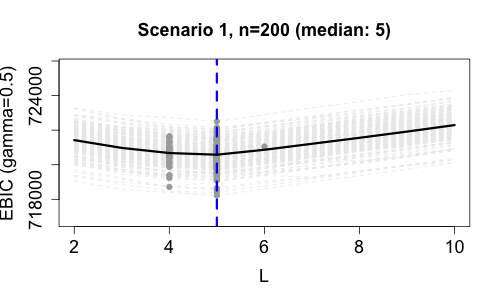}
    \includegraphics[width=0.3\linewidth]{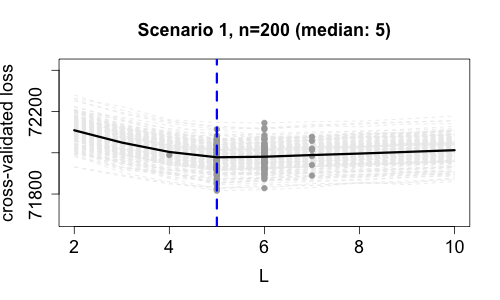}
    
    \includegraphics[width=0.3\linewidth]{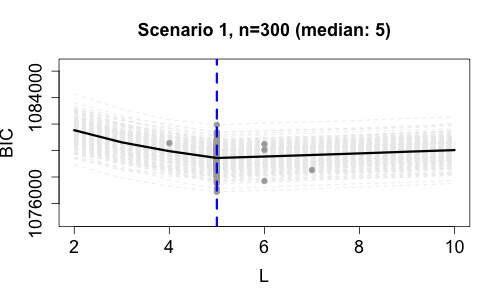}
    \includegraphics[width=0.3\linewidth]{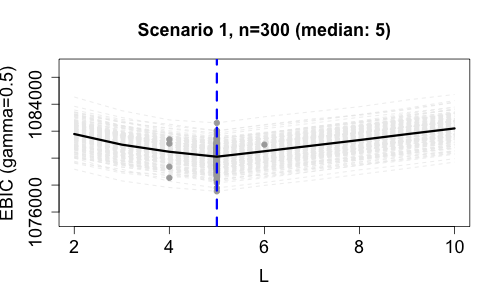}
    \includegraphics[width=0.3\linewidth]{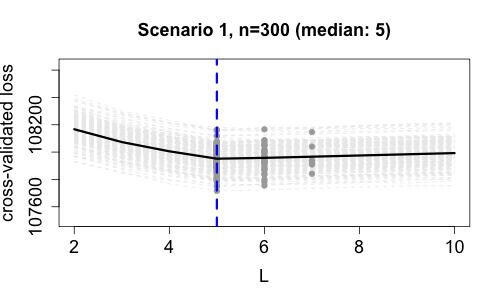}
    
    \includegraphics[width=0.3\linewidth]{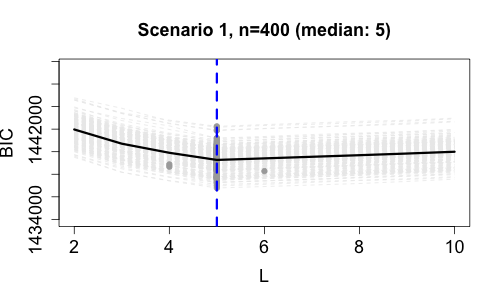}
    \includegraphics[width=0.3\linewidth]{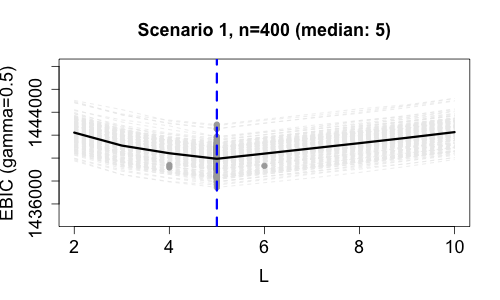}
    \includegraphics[width=0.3\linewidth]{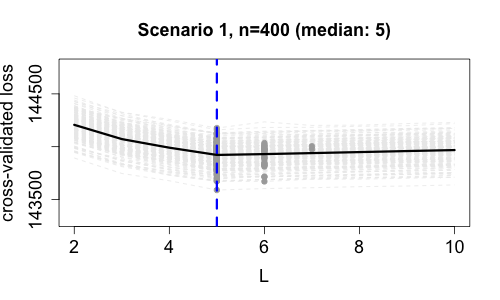}
    
    \includegraphics[width=0.3\linewidth]{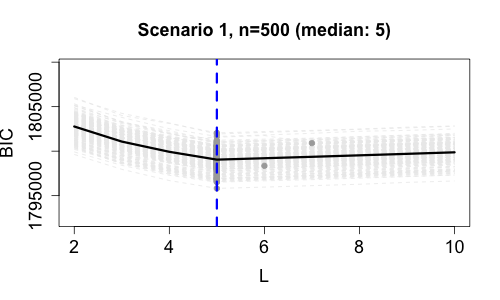}
    \includegraphics[width=0.3\linewidth]{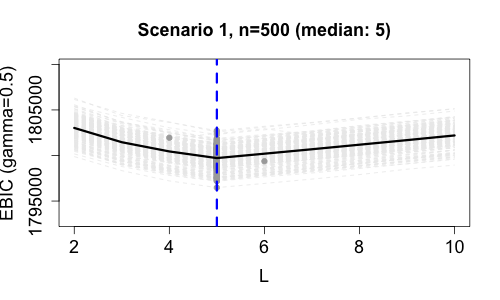}
    \includegraphics[width=0.3\linewidth]{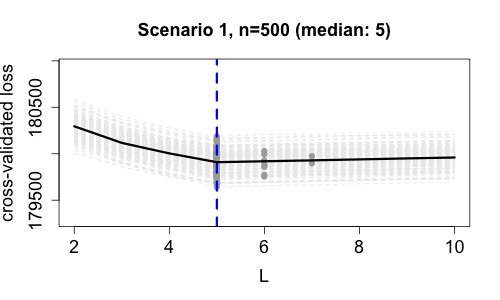}
    \caption{Evaluation of the BIC, EBIC $(\gamma=0.5)$, and 5-fold cross-validation in choosing $L$ in simulation studies, Scenario 1. Each dotted curve and gray dot denote the metric and its minimum for each simulated data, and the solid curve represents the pointwise average across simulations.}
    \label{fig:simulation2S1}
\end{figure}

\newpage

\begin{figure}[h]
    \centering
    \includegraphics[width=0.3\linewidth]{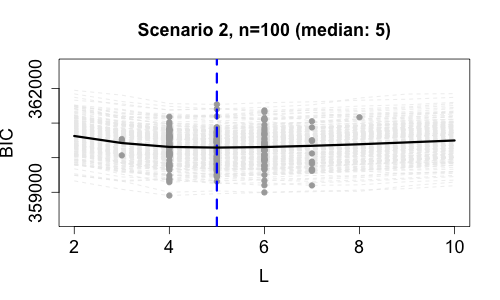}
    \includegraphics[width=0.3\linewidth]{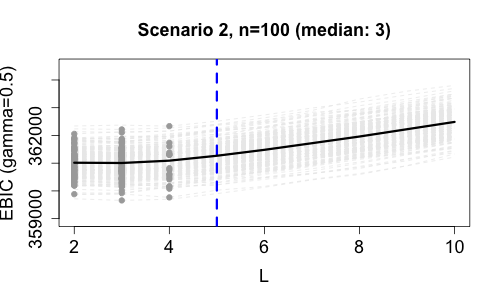}
    \includegraphics[width=0.3\linewidth]{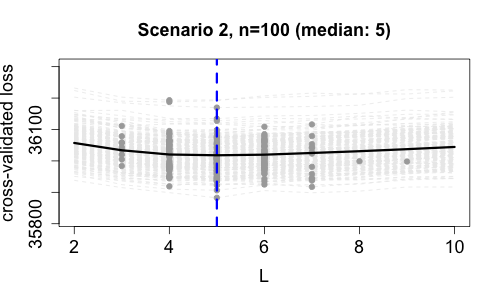}
    
    \includegraphics[width=0.3\linewidth]{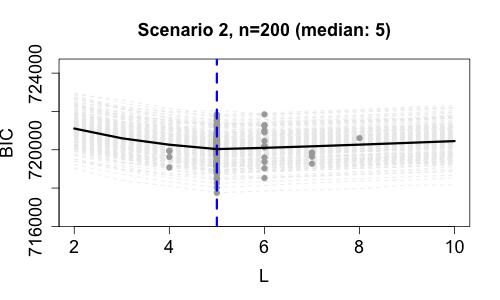}
    \includegraphics[width=0.3\linewidth]{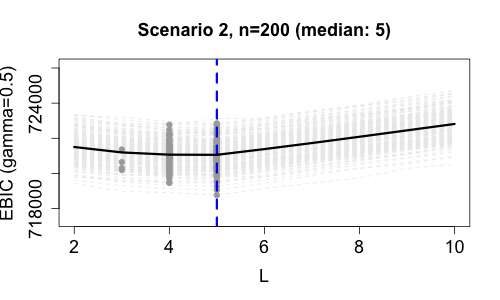}
    \includegraphics[width=0.3\linewidth]{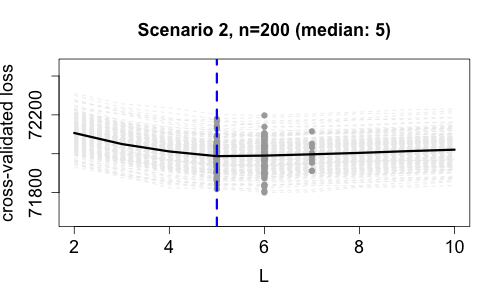}
    
    \includegraphics[width=0.3\linewidth]{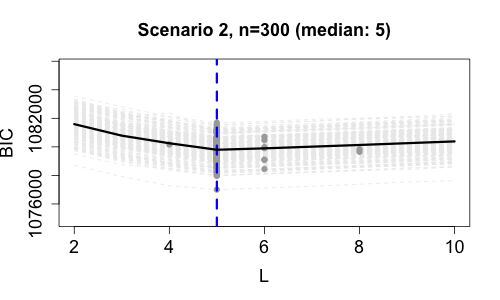}
    \includegraphics[width=0.3\linewidth]{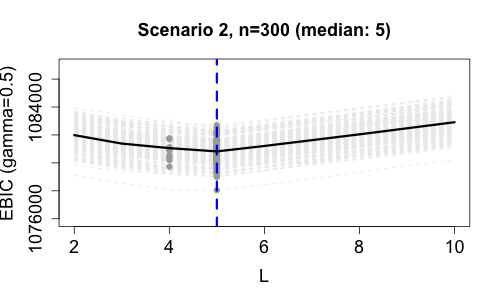}
    \includegraphics[width=0.3\linewidth]{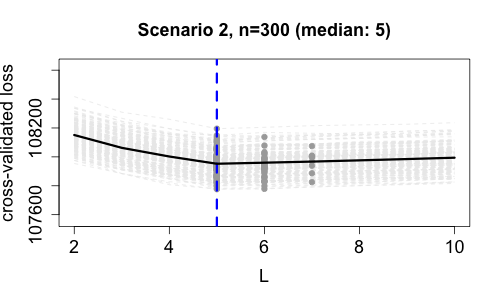}
    
    \includegraphics[width=0.3\linewidth]{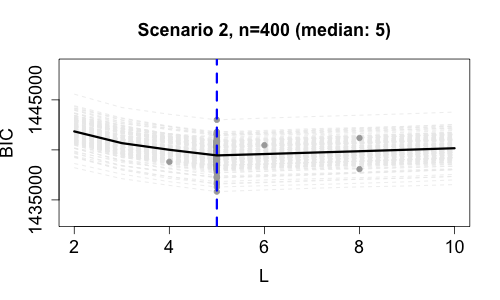}
    \includegraphics[width=0.3\linewidth]{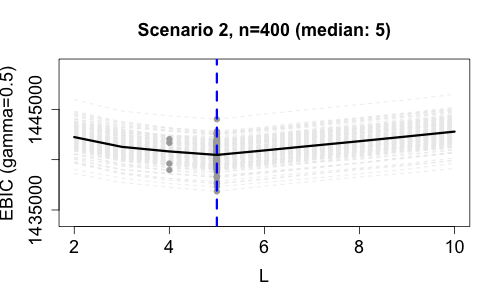}
    \includegraphics[width=0.3\linewidth]{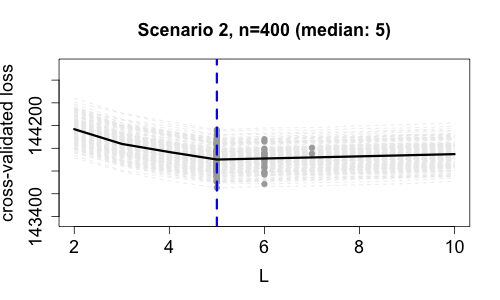}
    
    \includegraphics[width=0.3\linewidth]{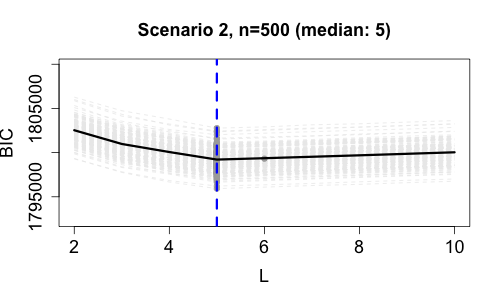}
    \includegraphics[width=0.3\linewidth]{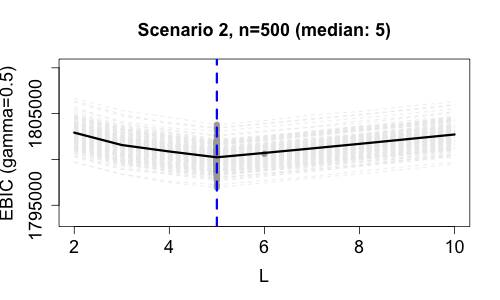}
    \includegraphics[width=0.3\linewidth]{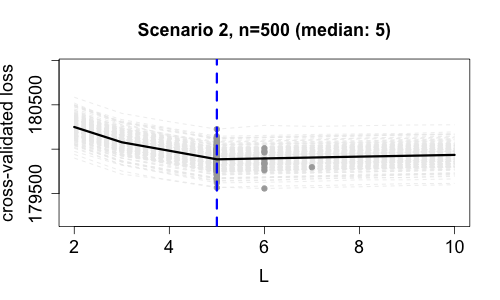}
    \caption{Evaluation of the BIC, EBIC $(\gamma=0.5)$, and 5-fold cross-validation in choosing $L$ in simulation studies, Scenario 2. Each dotted curve and gray dot denote the metric and its minimum for each simulated data, and the solid curve represents the pointwise average across simulations.}
    \label{fig:simulation2S2}
\end{figure}

\newpage
\section{Supplementary results for Section \ref{sec:data}}\label{sec:datasupp}

This section provides supplementary results for the analysis of the ABIDE data mapped onto the Fisher space. For each subject-specific connectivity matrix, where each edge is measured by the Pearson correlation, we applied a matrix-log transformation after adding a small constant, $1.0\times 10^{-10}$, to the diagonal entries, and considered the off-diagonal entries for analysis as in \citet{desai2025connectivity}. Because of the matrix-log transformation, no edge-wise centering was applied.

\begin{figure}[h]
    \centering
    \includegraphics[width=0.4\linewidth]{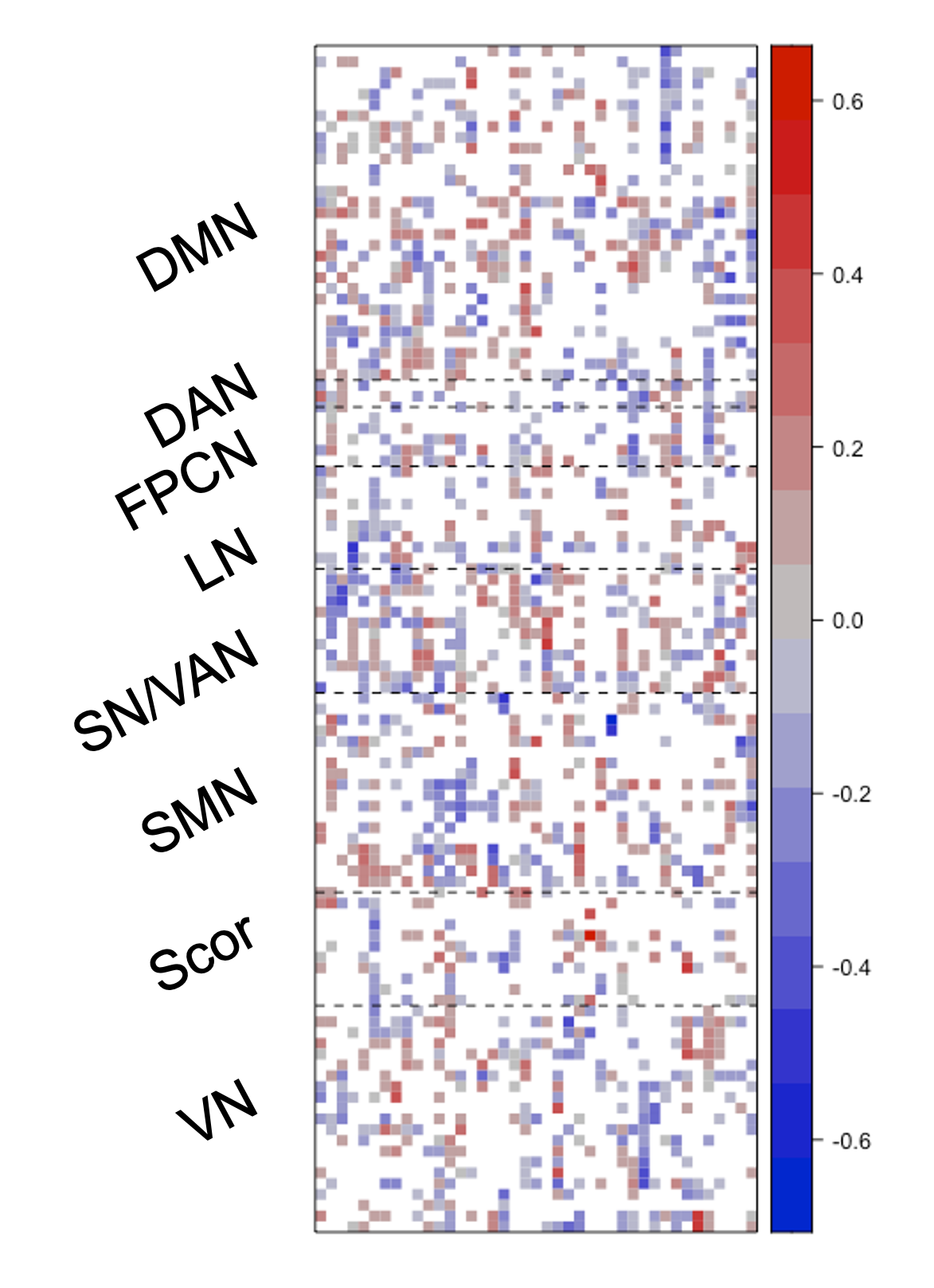}
    \caption{Estimated ${\bU}$ ($L=41$) using the training data ($n=591$) mapped onto the Fisher space, with the column orders obtained by hierarchical clustering.}
    \label{fig:Ulogm}
\end{figure}

The BIC selected \(L=41\) for the training data, and the estimated \(\bU\) is summarized in Figure \ref{fig:Ulogm}. Despite hierarchical clustering, the estimated \(\bU\) did not reveal noticeable network-specific patterns, although features in the DMN and SN/VAN appeared more relevant to covariate effects across \(L\) than features in the other networks.

\begin{figure}[t]
    \centering
    \includegraphics[width=0.4\linewidth]{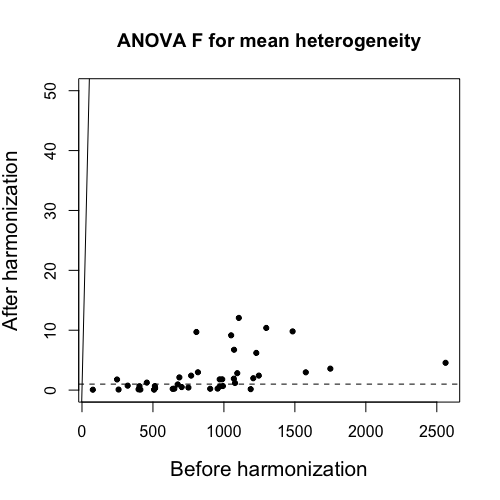}
    \includegraphics[width=0.4\linewidth]{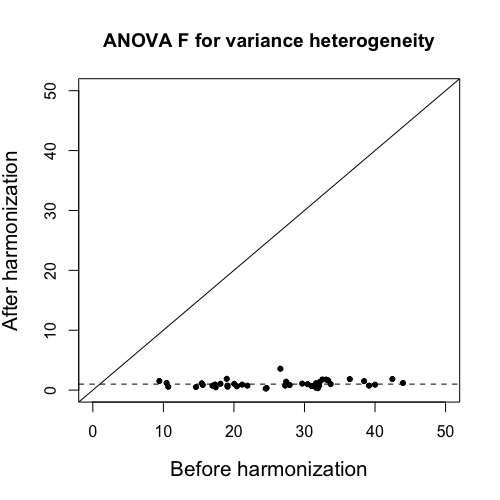}

    \includegraphics[width=0.4\linewidth]{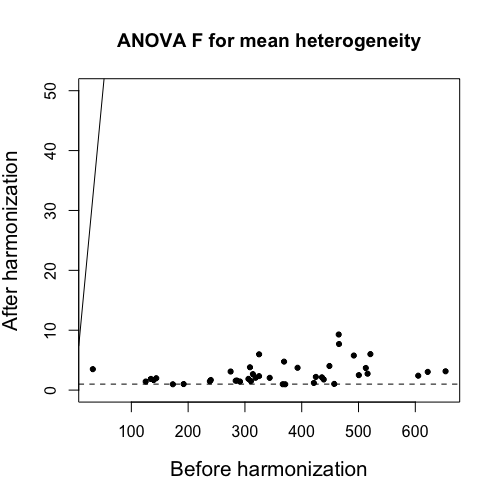}
    \includegraphics[width=0.4\linewidth]{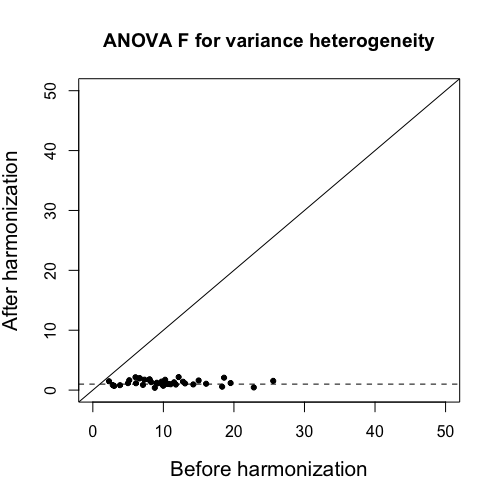}
        
    \caption{$F$ test statistic for measuring mean and variance heterogeneity across sites using the data mapped onto the Fisher space, using both training (first row) and test data (second row). The dotted horizontal line is the expected value of the null $F$ distribution (i.e., no site effects). The solid line is the 45 degree line.}
    \label{fig:ANOVAF_logm}
\end{figure}

We note, however, that the latent subject scores obtained in the test data by applying the training-data fit revealed much more noticeable site effects than those in the element-wise Fisher-transformed data analyzed in the main manuscript. We computed ANOVA $F$ test statistics to measure mean and variance heterogeneity in each subject score, and the results are summarized in Figure \ref{fig:ANOVAF_logm}. In the element-wise analysis, the SLACC-harmonized test dataset had median $F$ values of 1.10 and 1.09 for mean and variance heterogeneity, respectively, while the ComBat-harmonized test dataset had corresponding values of 1.22 and 1.15.

These supplementary results suggest that the matrix-log domain may provide a useful alternative representation for future extensions of SLACC when preserving correlation-matrix geometry is important.

\end{document}